\documentclass[preprint,12pt,year]{elsarticle}

\usepackage{amssymb,bm}
\usepackage{graphicx}
\usepackage{amsmath}
\usepackage{booktabs}
\usepackage{tabularray}
\UseTblrLibrary{booktabs}
\usepackage{tabularx}
\usepackage[export]{adjustbox}
\usepackage[colorlinks=true,
            linkcolor=blue,
            urlcolor=red,
            citecolor=blue]{hyperref}
\usepackage{natbib}
\usepackage{algorithm}
\usepackage{algpseudocode}
\journal{Nuclear Physics B}

\newcommand{\obs}[1]{\mathcal{H}\left(#1\right)}
\newcommand{\mean}[1]{\widetilde{\bm{\mu}}^{#1}}
\newcommand{\state}{\mathbf{x}}

\newcommand{\anomaly}[1]{\bm{\Lambda}^{#1}}
\newcommand{\covariance}[2]{\bm{\Sigma}^{#1,#2}}
\newcommand{\model}[2]{\mathcal{M}^{#1}\left(#2\right)}
\newcommand{\mb}[2]{\mathbf{#1}^{#2}}
\newcommand{\principal}[1]{\mb{x}{\bm{\chi}#1}}
\newcommand{\control}[1]{\mb{x}{\bm{\gamma}#1}}
\newcommand{\ancillary}[1]{\mb{x}{\bm{\eta}#1}}
\newcommand{\total}[1]{\mb{x}{\bm{\xi}#1}}

\begin{document}

\begin{frontmatter}



\title{Multi-fidelity Ensemble Kalman Filter algorithms enhanced by Convolutional Neural Networks} 

%
%


\author[inst1]{Tom Moussie} 
\author[inst1,inst2]{Paolo Errante}
\author[inst1]{Marcello Meldi}

\affiliation[inst1]{organization={Univ. Lille, CNRS, ONERA, Arts et Métiers ParisTech, Centrale Lille, UMR 9014- LMFL- Laboratoire de Mécanique des fluides de Lille},
            city={Kampé de Feriet},
            postcode={F-59000}, 
            state={Lille},
            country={France}}

\affiliation[inst2]{organization={L2EP, Laboratoire d'Electrotechnique et d'Electronique de Puissance de Lille},
            addressline={Avenue Henri Poincaré}, 
            city={Bâtiment ESPRIT},
            postcode={59655}, 
            state={Lille},
            country={France}}

\begin{abstract}
The present research work proposes advancement for Data Assimilation strategies using Convolutional Neural Networks (CNN). More precisely, multi-fidelity and multi-level algorithms for the Ensemble Kalman Filter are enhanced by CNN tools, with the objective to reduce the discrepancy in the prediction between ensemble realizations performed with different models. The proposed methodology is assessed via the analysis of the flow through a cascade of NACA\,0012 profiles for Reynolds $Re=1\,000$ and Mach $Ma=0.5$. Depending on the angle of attack $\alpha$, unsteady features of the flow can be observed. The results indicate that the usage of the CNN tools, which are trained using data from the DA procedure,  significantly augments the accuracy of the low-fidelity models with little augmentation in computational costs. It is shown that the usage of the CNN tools provides a faster convergence of the Data Assimilation algorithms, which leads to a significant gain in terms of computational resources required.  
\end{abstract}

\begin{graphicalabstract}
\end{graphicalabstract}

\begin{highlights}
\item Extended comparison of ensemble Data Assimilation techniques available in the literature using an \textit{online} CFD-DA architecture.
\item Development of a Machine Learning, CNN-based, predictive tool to enhance the accuracy of the low-fidelity models included in the multi-fidelity algorithms.
\item Integrating the Machine Learning tools into the Data Assimilation \textit{online} platform to perform \textit{on-the-fly} DA predictions.
\end{highlights}


\end{frontmatter}



\section{Introduction} \label{sec:intro}

The development of new numerical techniques and algorithms is a key step to obtain systematic technological advancement in real-life applications for numerous scientific disciplines. This is particularly true for studies in fluid mechanics, which are relevant to a large number cases such as the investigation of non linear, multiscale phenomena and their application to transport engineering, medical investigations, energy harvesting and urban settings. Therefore, progress in the accuracy of the predictive tools for this discipline will provide massive technological advancement for a wide spectrum of applications. For scale resolving tools such as Computational Fluid Dynamics (CFD) solvers \cite{Ferziger1996_springer}, the challenges to be faced towards this objective are twofold. On the one hand, numerical resources required to resolve all the dynamically active scales of the flow are prohibitive for high Reynolds numbers, which are observed for most applications in engineering and are characterized by a strongly unsteady behavior. One can envision the usage of closures to reduce such computational burdens \cite{Pope2000_cambridge}. This is particularly true for turbulence modeling. These strategies, which have been extensively investigated in approaches such as Reynolds Average Navier-Stokes (RANS) modeling \cite{Wilcox2006_DCW} and Large Eddy Simulation (LES) \cite{Sagaut2006_springer}, may exhibit relatively good accuracy but are extremely sensitive to the test case investigated. A second problem to be faced is that, even if the considered numerical model has perfect accuracy, comparisons with real applications may be precluded by lack of knowledge of initial\,/\,boundary conditions \cite{Meldi2020_jfm}. Because of the multiscale nature of high Reynolds flows, minimal variations in instantaneous values may lead to completely different structural organization of the observed eddies. For both cases, calibration of numerical ingredients of the algorithms using reference data such as samples from experiments is therefore a complex task to accomplish.

These challenges have been extensively studies in the last two decades using data-informed approaches. The studies reported in the literature include applications of Uncertainty Quantification (UQ) and Uncertainty Propagation (UP) \cite{Ghanem1991_springer,Xiao2019_pas}, Data Assimilation (DA) \cite{Daley1991_cambridge,Asch2016_SIAM} and Machine Learning \cite{Duraisamy2019_arfm,Brunton2020_arfm}. These approaches have provided guidelines of application for enhanced turbulence modelling \cite{Cherroud2025_jcp} and for the determination of suitable boundary conditions \cite{moussie_statistical_2024}. Among the family of approaches investigated, DA techniques show a high potential of application. They are complementary to UQ applications and they can be efficiently combined with ML approaches, when reference data used to calibrate such models is sparse in space and time \cite{Villiers2025_ftc}. This is the case for most industrial applications where the positioning of sensors in critical locations can be precluded. For studies in fluid mechanics, methods of variational DA have been extensively used for stationary / ensemble-averaged configurations, where the presence of a solution attractor permits to exploit the precision of such methods \cite{Foures2014_jfm,Mons2021_jfm}. For unsteady, time-resolved flows, ensemble approaches based on Bayesian theory have seen recent applications even for three-dimensional problems \cite{Labahn2019_pci,Villanueva2024_IJHFF}. These approaches notably include methods based on the Ensemble Kalman Filter (EnKF) \cite{evensen_sequential_1994}. One promising feature of such approaches is the possibility to combine predictive model with sparse data obtained in streaming from a real application, with the objective to control the physical configuration with a \textit{digital twin} \cite{Rasheed2020_IEEE}. Such complex application, which is at the core of objectives in Industry 4.0, is however problematic for application in fluid mechanics due to the aforementioned computational costs required. These costs are also increased because of the resources required to perform the DA calculations, which are usually even larger that the ones required for time advancement of the CFD solver. Present applications are therefore restricted to physical flow phenomena characterized by slow turnover times, such as astronomic flows.

In order to obtain technological advancement in this field, strategies to reduce computational resources of DA while keeping the same degree of accuracy have been proposed. Strategies including localization and inflation are well documented and effective to this purpose \cite{Asch2016_SIAM}. For ensemble approaches such as the EnKF, another possibility is to rely on multi-fidelity techniques. These approaches usually combine few realizations of high-precision numerical models with a large number of low-precision runs, in order to optimize accuracy, computational cost and robustness. Among the proposals reported in the literature, \citet{popov_multifidelity_2021} developed a multi-fidelity EnKF (MFEnKF), which uses heterogeneous model realizations and combines their results within a control variate algorithm. Following a different route, Moldovan et al. \cite{moldovan_multigridensemble_2021} relied on multi-level / multi-grid capabilities of numerical solvers to obtain flow representations with different grid resolution. Within this algorithm, they coupled the classical DA analysis phase (referred to \textit{outer loop}) with a second DA update, where data from the high-fidelity model is used to calibrate some corrective term included in the low-fidelity solver. The latter is referred to as \textit{inner loop}.

In the present work, a new technique is proposed to enhance multi-fidelity ensemble DA approaches using ML techniques. More precisely, results from DA application are used to train a corrective term for the low-fidelity model, improving its precision. This corrective term is obtained using Convolutional Neural Networks (CNN) \cite{zhao_review_2024}, which are fed with instantaneous and complete flow snapshots provided by the DA algorithm. An interesting feature of this strategy is that, thanks to the possibility of the DA tool to generate complete sets of data over a large parametric space of investigation, the convergence of the CNN tools is significantly more rapid and robust. This analysis envisions future applications where streaming data from limited sensors of a physical application are going to be fed to a multi-fidelity DA algorithm. This architecture will simultaneously improve the numerical models with the data instantaneously available and, in a closed loop, it will provide efficient control strategies for physical / digital twin architecture. The proposed algorithm is here investigated via the analysis of a cascade of NACA\,0012 profiles for $Re=1\,000$ and $Ma=0.5$. For this configuration unsteady features can be observed, requiring therefore a complex CNN corrective term able to take into account time variations of the flow.

The article is structured as follows. In \autoref{sec:numerical_tools}, numerical ingredients about the CFD solvers and the test case of investigation is presented. In \autoref{sec:data_driven_methodologies}, the data-driven methodologies are introduced. In \autoref{sec:data_driven_optimization_inlet_BC}, state-of-the-art EnKF as well as multi-fidelity DA investigations are performed and discussed.  \autoref{sec:EnKF_with_CNN} discusses the rationale of the coupled DA-ML methodology, then the enhancement of the multi-fidelity algorithms by the CNN tools. Finally, in \autoref{sec:conclusion} the concluding remarks are drawn and future perspectives are discussed. 

\section{Numerical tools and test case} \label{sec:numerical_tools}

\subsection{Governing equations and numerical solvers}

The numerical solvers employed in this study are based on the Finite-Volume discretization \cite{Ferziger1996_springer} of the compressible Navier-Stokes equations for Newtonian fluids. The set of equations considered is:
\begin{equation} \scriptsize
    \frac{\partial \rho}{\partial t} + \nabla \cdot (\rho \mathbf{u}) = 0 \label{eq:continuity}
\end{equation}
\begin{equation} \scriptsize
    \frac{ \partial \rho \mathbf{u}}{\partial t} + \nabla \cdot \left( \rho \mathbf{u} \otimes \mathbf{u} \right) = -\nabla p + \nabla \cdot \mu \left( \nabla \otimes \mathbf{u} + \left( \nabla \otimes \mathbf{u}\right)^\top - \frac{2}{3} \left(\nabla \cdot \mathbf{u}\right) \cdot \delta_{ij} \right) \label{eq:momentum}
\end{equation}
\begin{equation} \scriptsize
    \frac{\partial \rho E}{\partial t} + \nabla \left( \rho E+p \right) \mathbf{u} = \nabla \cdot \left[ \mu \left( \nabla \otimes \mathbf{u} + \left( \nabla \otimes \mathbf{u}\right)^\top - \frac{2}{3} \left(\nabla \cdot \mathbf{u}\right) \cdot \delta_{ij} \right) \cdot \mathbf{u} - \kappa \nabla T \right] \label{eq:energy}
\end{equation}

where $\mathbf{u}$ is the velocity field, $p$ is the pressure, $\rho$ is the density, $E$ is the total energy, $T$ is the temperature $\mu$ is the dynamic viscosity and $\kappa$ is the thermal conductivity. $\delta_{ij}$ is the Kronecker symbol. \autoref{eq:continuity}, \autoref{eq:momentum} and \autoref{eq:energy} represent mass, momentum and energy conservation, respectively.

Numerical simulations are performed via discretization of the dynamic equations using the C++ open-source tool OpenFOAM. This library includes a number of solvers based on the Finite Volumes (FV) discretization. Differences among solvers are due to the features of the physical configuration to be investigated (turbulence, compressibility effects, heat transfer...) which usually require specific algorithms to grant maximized efficiency. In this work the solver \texttt{rhoPimpleFoam} is used. This solver is based on the Pimple algorithm \cite{caretto_simple_1973, ISSA198666} and it is tailored to the analysis of unsteady, compressible flows. OpenFOAM offers multiple schemes for the discretization of time and space derivatives. A second-order backward scheme is used to perform time advancement. Concerning the discretization in space, a native second-order centered scheme is used for both advective and diffusive terms.

\subsection{Flow around a cascade of NACA\,0012 profiles}
\begin{figure}[h!]
    \centering
    \includegraphics[width=0.99\linewidth]{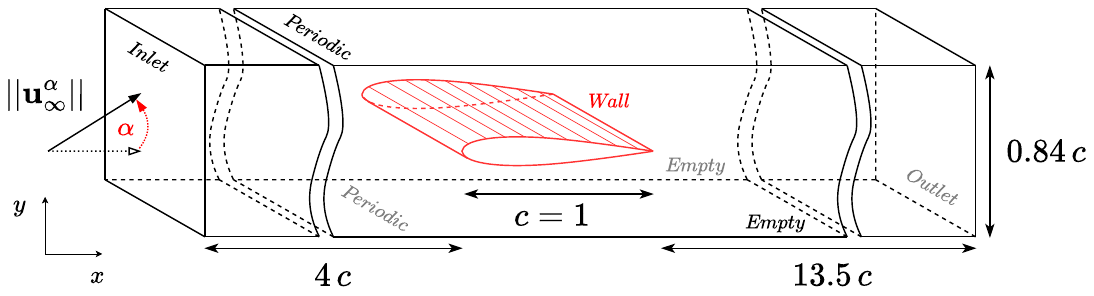}
    \caption{Computational domain and boundary conditions.}
    \label{fig:domain_and_BC}
\end{figure}

The test case of investigation selected for the present analysis is the two-dimensional flow around a cascade of NACA\,0012 airfoils \cite{YILBAS1998143}. This case, which is shown in \autoref{fig:domain_and_BC}, is characterized by a simple geometry permitting for rapid CFD investigations using a limited number of mesh elements. However, physical features such as boundary layer development, separation and stall can be unambiguously studied. 
The test case is now described. The profile is aligned along the $x$ direction of the coordinate system and $y$ is the normal direction. The origin of the system is set on the leading edge of the profile. The NACA\,0012 airfoil has a maximum thickness equal to 12\,\% of the chord length. The wing profile is obtained from \autoref{eq:naca0012} \citep{ladson_computer_1996}:

\begin{equation}
    \pm \frac{y^\prime}{c} = \frac{t/c}{0.2} \left[a_0 \left(\frac{x}{c} \right)^{\frac{1}{2}} + a_1 \left(\frac{x}{c} \right) + a_2 \left( \frac{x}{c}\right)^2 + a_3\left( \frac{x}{c} \right)^3 + a_4 \left( \frac{x}{c} \right)^4 \right] \label{eq:naca0012}
\end{equation}

where $y^\prime$ is the airfoil thickness distribution along $x \in [0,c]$. Here $c$ is the chord length and $t/c=0.12$ is the maximum thickness-to-chord ratio. The constant values $a_i, \, i \in [0,4]$ are given in \autoref{eq:a_i constants}:
\begin{align}
    \begin{split}
    a_0 = 0.2969 &, \quad a_1 = -0.1260 ,\\
    a_2 = -0.3516 &, \quad a_3 = 0.2843, \\
    a_4 = -0.1015 &
    \label{eq:a_i constants}
    \end{split}
\end{align}

The numerical domain has size of $x \times y = [-4\,c, \, 14.5\,c] \times [-0.48\,c, \, 0.48\,c]$ and is shown in \autoref{eq:naca0012}. At the inlet, a constant velocity magnitude $\Vert \mathbf{u}_\infty \Vert$ is set so that the Reynolds number $Re = \rho \Vert \mathbf{u}_\infty \Vert c / \mu = 1\,000$ and the upstream Mach number is $Ma_\infty = \Vert \mathbf{u}_\infty \Vert / a_\infty = 0.5$, where $a_\infty$ is the speed of sound. $a_\infty$ is defined as $a_\infty=\sqrt{\gamma \, r \, T_\infty}$, where 
$T_\infty$ is the inlet temperature, 
$r$ is the specific gas constant and $\gamma$ the heat capacity ratio. For these conditions the flow is laminar and it shows two-dimensional features. The incidence of the profiles with respect to the upstream flow is controlled changing the angle of attack $\alpha$ (see \autoref{fig:domain_and_BC}). This parameter is varied via manipulation of the velocity field imposed at the inlet, without any modification to the computational grid: 
\begin{equation}
    \begin{bmatrix}
        u_{\infty,x} \\
        u_{\infty,y}
    \end{bmatrix}
    = \vert \vert \mathbf{u}_\infty \vert \vert \cdot
    \begin{bmatrix}
        \cos(\alpha) \\
        \sin(\alpha)
    \end{bmatrix}
    \label{eq:VelInlet}
\end{equation}

\begin{figure}[h!]
    \centering
    \begin{tabular}{cc}
        \includegraphics[width=0.45\textwidth]{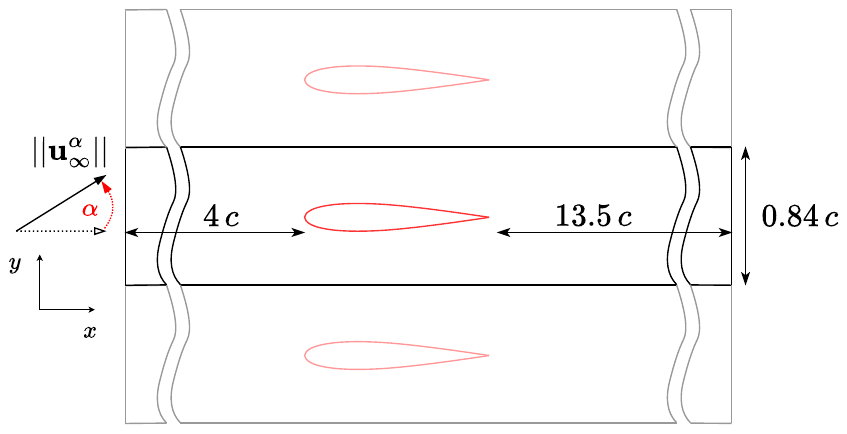} & \includegraphics[width=0.45\textwidth]{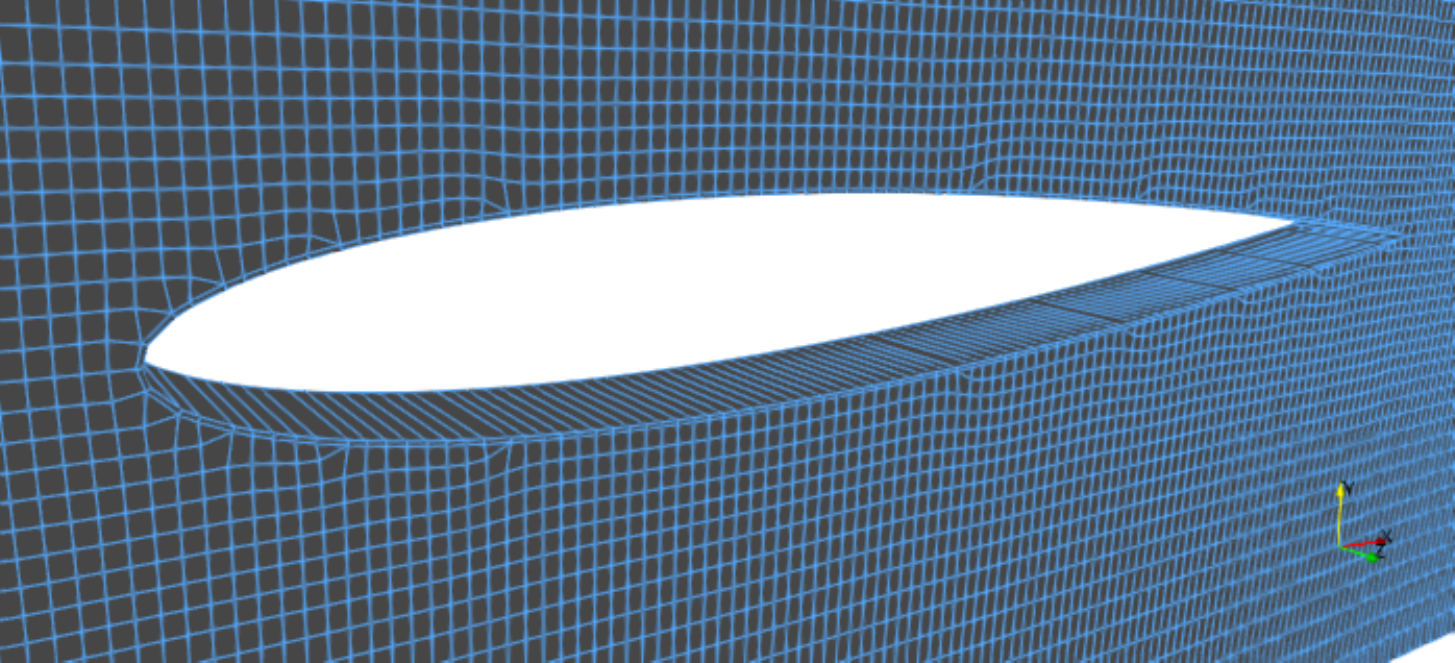} \\
        {\small (a)} & {\small (b)}
    \end{tabular}
    \caption{(a) Two-dimensional numerical domain and variation of the angle of attack. (b) Example of grid generated by OpenFOAM's tool \texttt{snappyHexMesh}.}
    \label{fig:domain_2D}
\end{figure}

A qualitative representation is shown in \autoref{fig:domain_2D} (a). As it will be shown, variations of $\alpha$ are responsible for the transition from a steady flow (low $\alpha$) to an unsteady regime (higher values of $\alpha$). Periodic boundary conditions are imposed on the top and bottom surface of the domain. The spacing in between air-foil is chosen so to obtain a solidity parameter of $s=0.84\,c$. This value has been chosen according to previous works in the literature \cite{YILBAS1998143}. Because of the FV discretization used, OpenFOAM requires a three-dimensional grid even for two-dimensional calculations. In this case, the grid exhibits only one element in the spanwise direction $z$ and empty (i.e. no-flux) boundary conditions are imposed on the lateral sides. At the outlet, a non-reflecting, mass conserving boundary condition is applied. This OpenFOAM-native boundary condition, named \textit{inletOutlet}, is also set to prevent reverse flows. A static pressure Dirichlet condition 
is imposed at the outlet boundary. Since small variations of the temperature $T$ are observed for this test case, the dynamic viscosity $\mu$ and the thermal conductivity $\kappa$ are considered to be constant. The performance of the OpenFOAM code was assessed with the preliminary analysis of the flow around a NACA\,0012 profile in an open domain. The results obtained are in agreement with findings reported in the literature for the same test case \cite{kurtulus_unsteady_2015}.

\subsection{Grid convergence study and preliminary simulations}
\label{sec::2.3}

\begin{figure}[h!]
    \centering
    \begin{tabular}{cc}
      \includegraphics[width=0.45\linewidth]{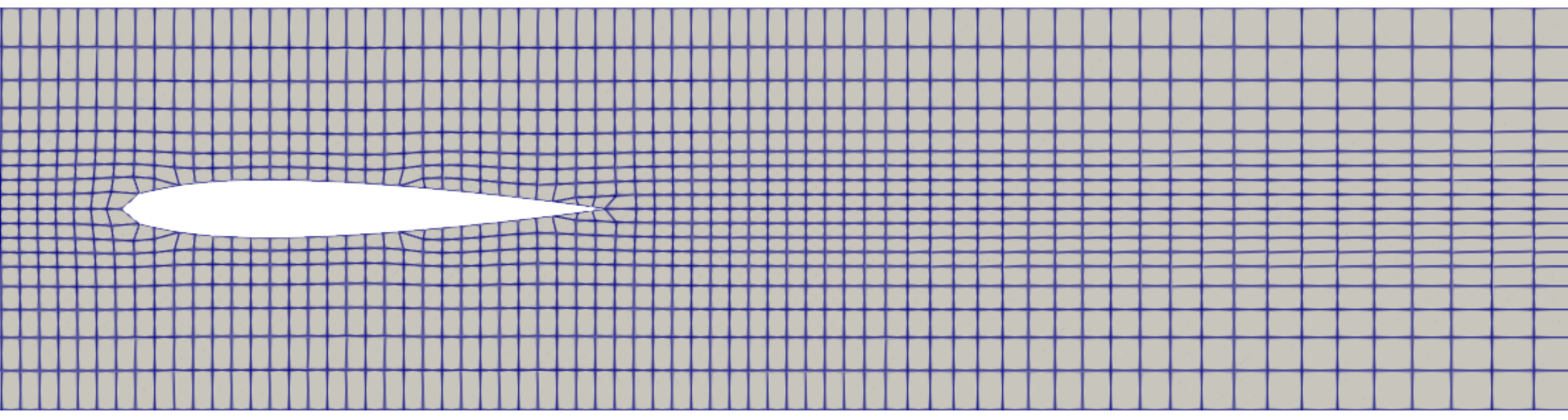}   & \includegraphics[width=0.45\linewidth]{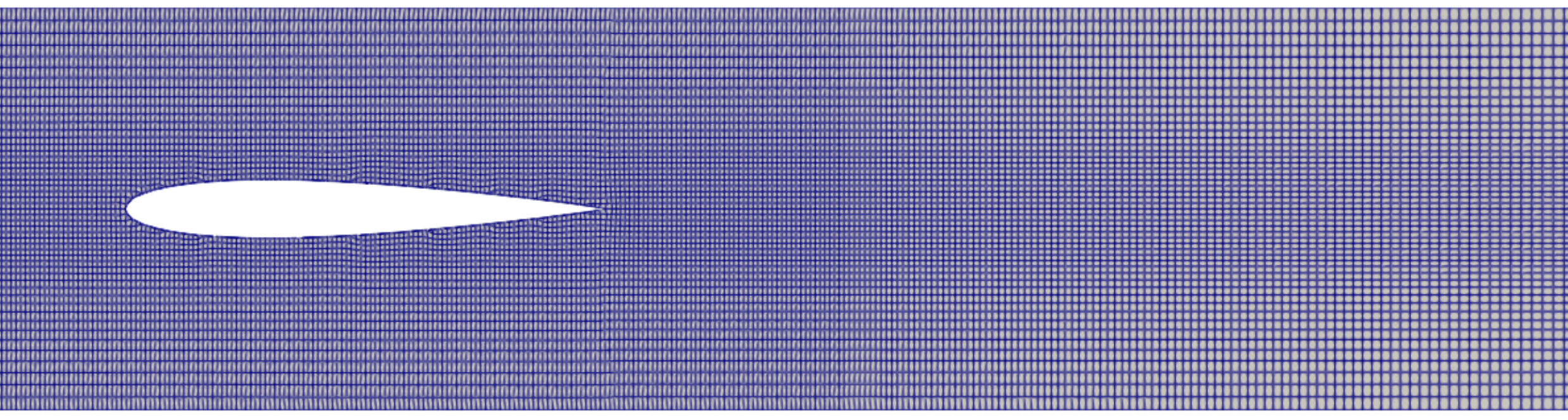} \\
      {\small (a)} & {\small (b)}   \\[0.25cm]
      \includegraphics[width=0.45\linewidth]{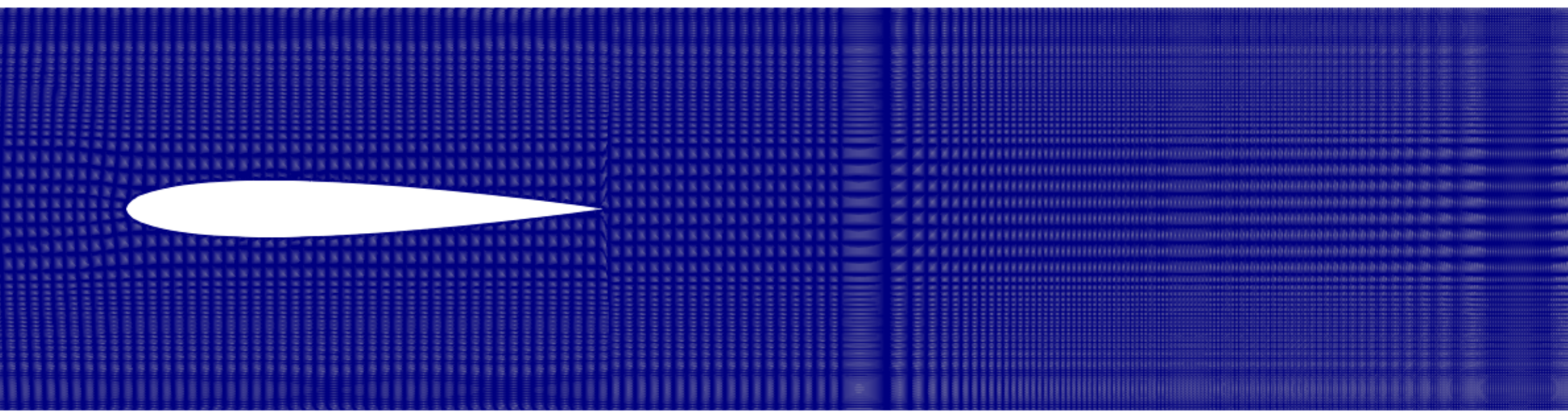}   &   \\
      {\small (c)}
    \end{tabular}
    
    \caption{Qualitative representation of the grid resolution in the wake area. (a) Coarse grid; (b) Fine grid; (c) Reference grid.}
    \label{fig:cascade_meshes}
\end{figure}

Initial simulations were performed to assess the sensitivity of the results to the grid resolution. To this purpose, three grids here referred as Reference grid, Fine grid and Coarse grid have been used. They are created with the OpenFOAM native utility \texttt{snappyHexMesh} and grid details are given in \autoref{tab:grid_refine_desc}. A visualization around the profile is shown in \autoref{fig:domain_2D} (b) while a qualitative representation of the grid resolution in the wake region is shown in \autoref{fig:cascade_meshes}. For the Reference grid and the Fine grid, layers of elements in the proximity of the immersed profile are generated. This choice permits to obtain an improved resolution of near-wall dynamics.  For the Fine grid displayed on \autoref{fig:cascade_meshes}\,(b), a single layer is set with a size in the wall normal direction of $3\cdot10^{-3}\,c$. The Reference grid shown on \autoref{fig:cascade_meshes}\,(c) is created using ten layers around the profile. The closest to the wall has a thickness of $7\cdot10^{-5}\,c$. Moving away from the profile, the thickness of the layers increases with an expansion ratio equal to $1.5$. 

\begin{table}[h!]
    \centering
    \footnotesize
    \begin{tabular}{lccc} \toprule[1.5pt]
        & \textbf{Elements} & \textbf{Nodes around airfoil} & \textbf{Near-wall layers} \\ \cmidrule{2-4}
       Coarse grid & $2\,600$  & $52$ & $0$ \\
       Fine grid & $30\,000$ & $202$ & $1$ \\
       Reference grid & $540\,000$ & $1\,616$ & $10$ \\\bottomrule[1.5pt]
    \end{tabular}
    \caption{Details about the computational grids used for the analysis of the NACA\,0012 cascade configuration.}
    \label{tab:grid_refine_desc}
\end{table}

\begin{figure}[h!]
    \centering
    \begin{tabular}{cc}
        \includegraphics[width=0.45\textwidth]{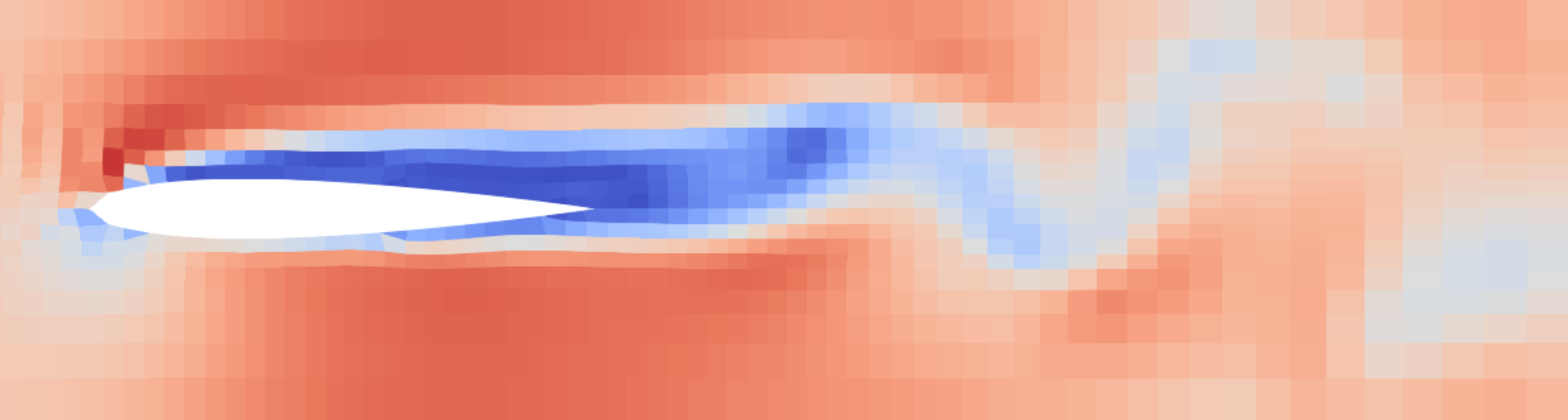}  & \includegraphics[width=0.45\textwidth]{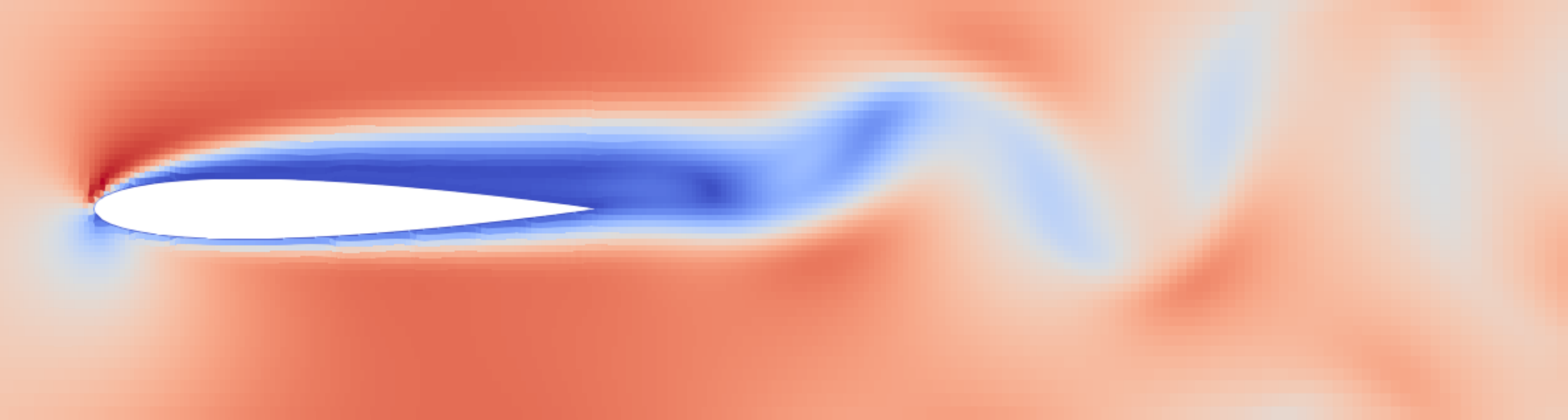}  \\
       {\small (a)} & {\small (b)} \\[0.25cm]
        \includegraphics[width=0.45\textwidth]{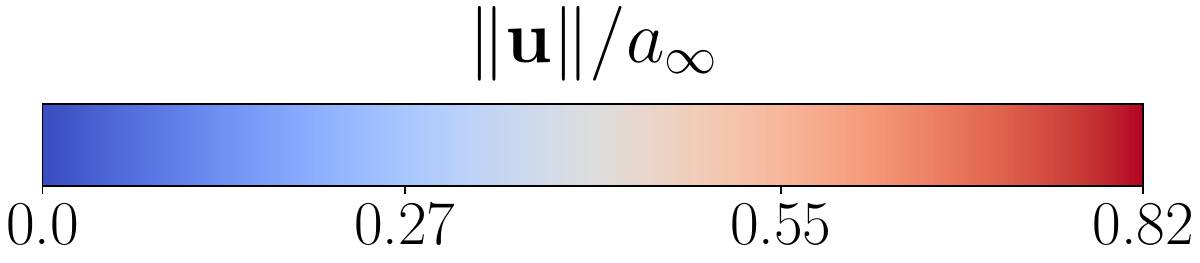} & \includegraphics[width=0.45\textwidth]{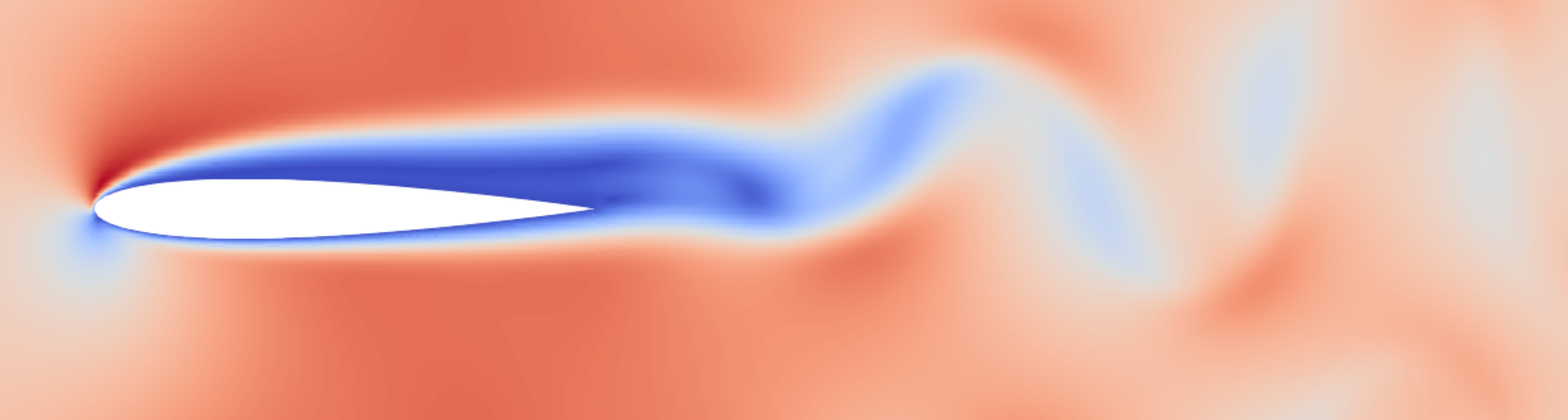} \\
        & {\small (c)}
    \end{tabular}
    \caption{Isocontours of the normalized instantaneous velocity magnitude for an angle of attack $\alpha=25^\circ$ at $t=25$. (a) Coarse grid; (b) Fine grid; (c) Reference grid.}
    \label{fig:AoA25_meshes}
\end{figure}

\begin{figure}[h!]
    \centering
    \begin{tabular}{c}
    \includegraphics[width=0.95\linewidth]{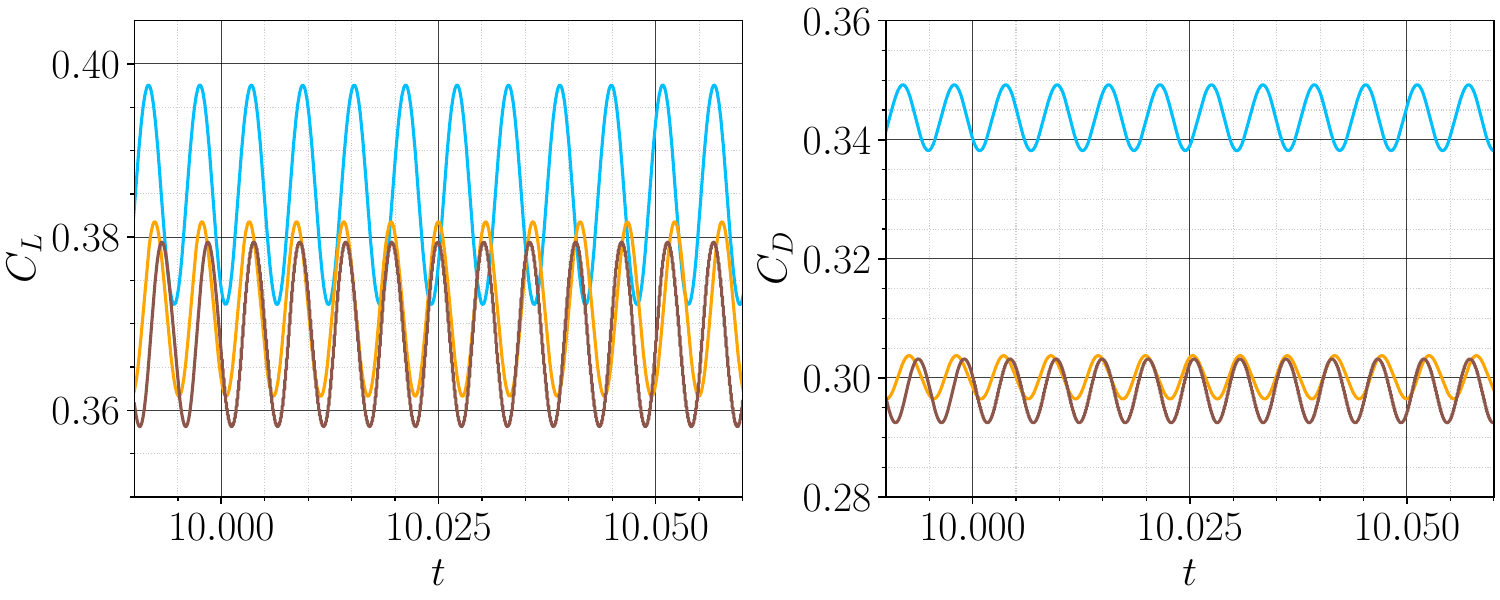} \\
    \includegraphics[width=0.9\linewidth]{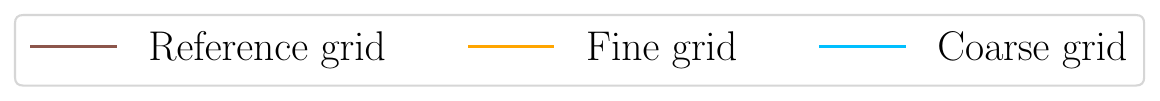} \\
    \end{tabular}
    \caption{Instantaneous lift and drag coefficients obtained by simulations using the Reference grid, the Fine grid and the Coarse grid for the range $t= [10,10.05]$. The value of the angle of attack is here set to $\alpha=25^\circ$.}
    \label{fig:lift_drag_instantaneous}
\end{figure}

Preliminary simulations are run using the three grids previously presented. Multiple runs are performed for different angle of attacks $\alpha = \left[0,8,12,20,25\right]^\circ$ changing the velocity boundary condition at the inlet according to \autoref{eq:VelInlet}. 
The simulations run using the Reference grid employ a time step value of $\delta t = 10^{-5}$. This choice permits to comply with a maximum local Courant number lower than $2$. On the other hand, simulations using the Fine grid and the Coarse grid are performed using $\delta t = 10^{-4}$. This choice provides a maximum local Courant number of approximately $3.5$ and $1$, respectively. All simulations are performed for a total physical time $T = 25\,t = 4\,250\,t_A$, where $t_A = c / \vert \vert \mathbf{u}_\infty \vert \vert$ is the characteristic advective time. The results, including the lift coefficient $C_L$, the drag coefficient $C_D$ and the Strouhal number, are presented in \autoref{tab:preliminary_results}. The Strouhal number, indicated as $St$, is calculated as \cite{kurtulus_unsteady_2015}:
\begin{equation}
    St = \frac{f\cdot c}{\vert \vert \mathbf{u}_\infty \vert \vert} \label{eq:strouhal}
\end{equation}
where $f$ is the characteristic frequency derived from the Fast-Fourier-Transform (FFT) analysis of the time-evolution of the lift coefficient $C_L$. Statistics were computed using the signals for $C_L$ and $C_D$ over the time window $[5,\,25]\,t$. The initial transient period $[0, 5]\,t$ was excluded from the statistical analysis in order to exclude effects of initial conditions.
The simulations performed indicate that steady flows configurations are observed for $\alpha \leq 20^\circ$, while unsteady features are observed for $\alpha=25^\circ$. The solidity $s$ of the cascade configuration is responsible for the delay of the transition towards the unsteady regime, which is observed for smaller angles of attack for a single profile. For the steady regimes, small differences are observed for the predictions of the time-averaged quantities $\overline{C_L}$ and $\overline{C_D}$ using the Refined grid and the Fine grid. Larger discrepancies, up to $16\,\%$, are instead obtained comparing results from the Reference grid with the Coarse grid. 
\begin{table}[h!]
\small
\centering
\begin{tabular}{lccccccccc}
\toprule[1.5pt]
       & \multicolumn{3}{c}{\textbf{Reference grid}} & \multicolumn{3}{c}{\textbf{Fine grid}} & \multicolumn{3}{c}{\textbf{Coarse grid}} \\ \cmidrule{2-10}
$\alpha$ [deg]  & $\overline{C_L}$ & $\overline{C_D}$  & $St$   & $\overline{C_L}$ & $\overline{C_D}$  & $St$  & $\overline{C_L}$ & $\overline{C_D}$  & $St$  \\ \midrule
0   & $\sim 0$  & 0.19  & n/a    & $\sim 0$       & 0.20  & n/a  & $\sim 0$        & 0.21   & n/a   \\
8   & 0.15      & 0.20  & n/a    & 0.15  & 0.20  & n/a  & 0.16   & 0.21   & n/a   \\
12  & 0.22      & 0.21  & n/a    & 0.22  & 0.21  & n/a  & 0.23    & 0.24   &  n/a\\
20  & 0.33      & 0.25  & n/a    & 0.33  & 0.26  & n/a   & 0.34   & 0.29   & n/a    \\
25  & 0.37      & 0.30  & 1.11   & 0.37  & 0.30  & 1.08  & 0.39   & 0.34    & 0.99   \\ \bottomrule[1.5pt]
\end{tabular}
\caption{Results of the preliminary simulations using the three grids for the five angle of attack investigated. The time-averaged lift coefficient $\overline{C_L}$, the time-averaged drag coefficient $\overline{C_D}$ and the Strouhal number are reported.}
\label{tab:preliminary_results}
\end{table}
Simulations using an angle of attack of $\alpha = 25^\circ$ exhibit a wake characterized by vortex shedding. A qualitative illustration of the downstream flow patterns at this angle of attack is presented in \autoref{fig:AoA25_meshes}. While the wake appears to be similar for the three grids used, $St=1.11$ for the Reference grid, $St=1.08$ for the Fine grid and $St=0.99$ for the Coarse grid. The analysis of the standard deviation of the lift and drag coefficients confirms that grid resolution affects the flow prediction. 
For the lift coefficient, the simulation using the Reference Grid yields values of $C_L^\prime = 0.0075$, while the Fine Grid provides to $C_L^\prime = 0.0071$ and the Coarse Grid $C_L^\prime = 0.009$. For the drag coefficient $C_D^\prime = 0.0037$ for the Reference Grid, $C_D^\prime = 0.0026$ for the Fine Grid and $C_D^\prime = 0.0039$ for the Coarse Grid. While noticeable, these differences are however lower in magnitude than the discrepancies observed for the time-averaged values in \autoref{tab:preliminary_results}. A confirmation is obtained by the observation of the time history of $C_L$ and $C_D$ over 13 shedding cycles, which is reported in \autoref{fig:lift_drag_instantaneous}. 

\begin{figure}[h!]
    \centering
    \begin{tabular}{cc}
      \includegraphics[width=0.45\linewidth]{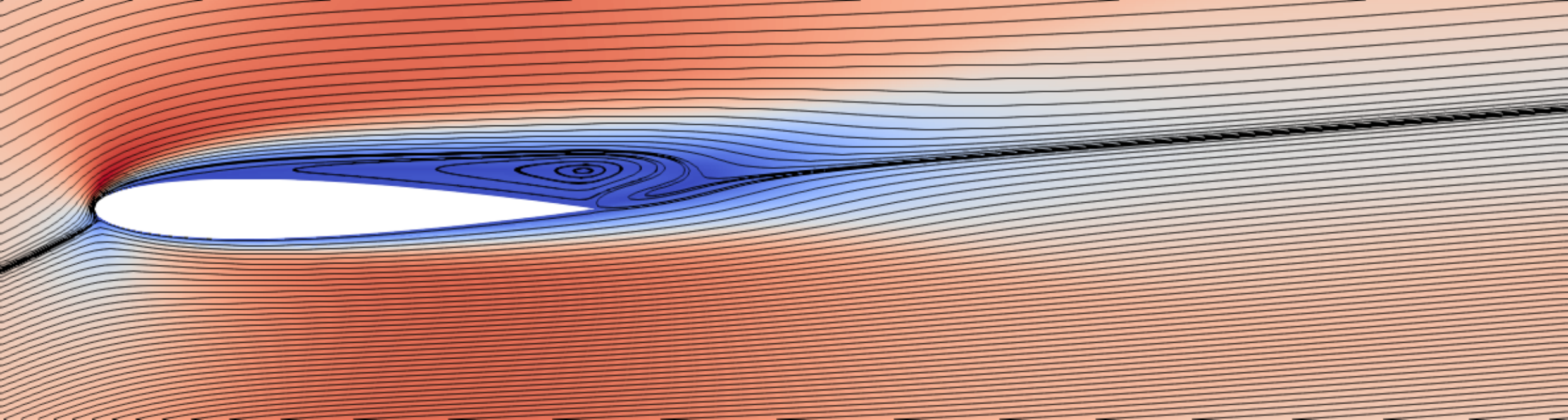}  & \includegraphics[width=0.45\linewidth]{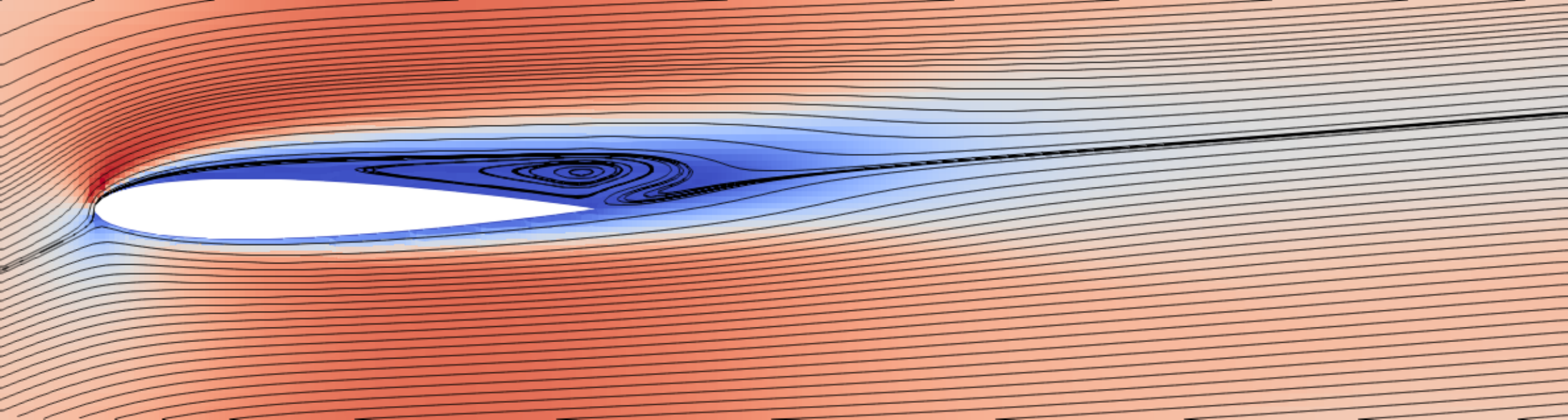} \\
      {\small (a)} & {\small (b)} \\[0.25cm]
        \includegraphics[width=0.45\linewidth]{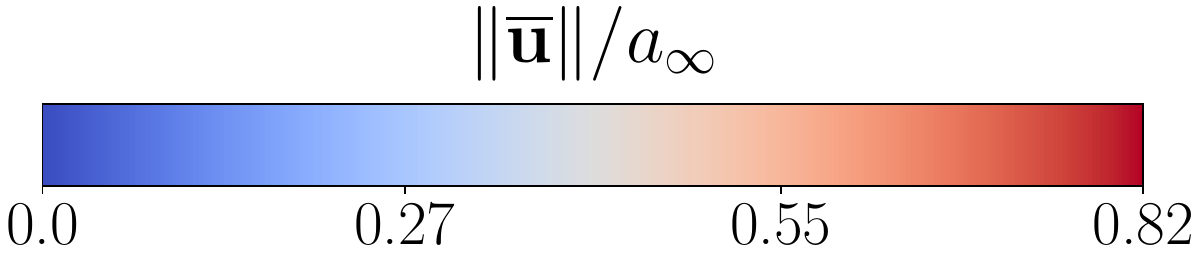} & \includegraphics[width=0.45\linewidth]{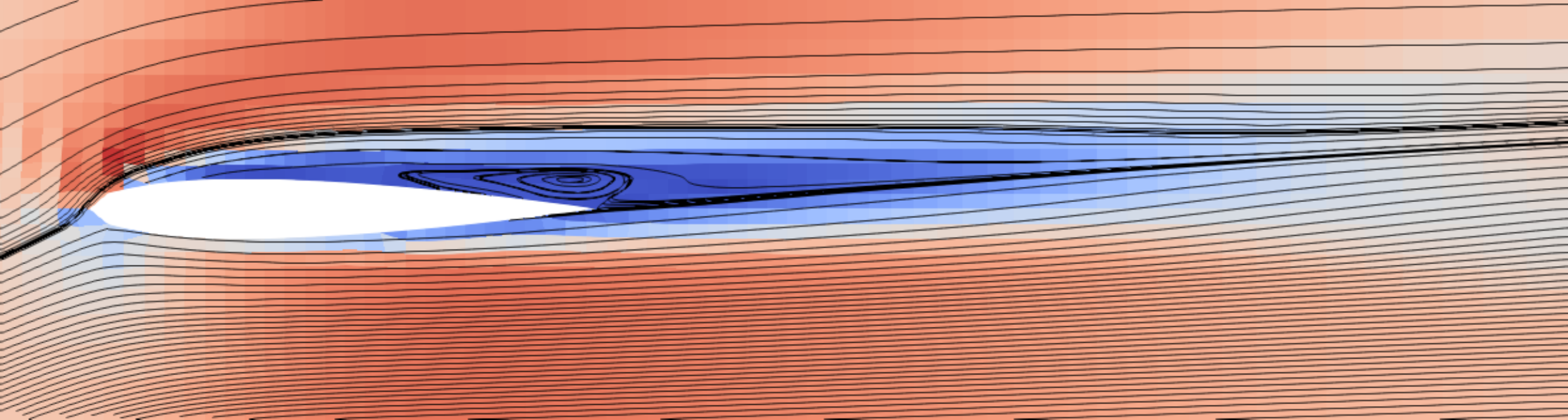} \\
        &   {\small (c)}
    \end{tabular}
    \caption{Mean velocity streamlines for an angle of attack $\alpha=25^\circ$ at $t=25$. (a) Reference mesh; (b) Fine mesh; (c) Coarse mesh. Background coloring produced with normalized mean velocity magnitude $\Vert \overline{\mathbf{u}} \Vert / a_\infty$ isocontours.}
    \label{fig:streamlines_UMean}
\end{figure}

The time-averaged flow field is now investigated for the simulations with $\alpha=25^\circ$. The iso-contours of the normalized time-averaged velocity magnitude $\Vert \overline{\mathbf{u}} \Vert / a_\infty$ is shown in \autoref{fig:streamlines_UMean} with streamlines representing the recirculation region due to separation of the boundary layer. While results obtained from the Reference grid and the Fine grid are qualitatively similar, one can see that significantly different flow features are obtained using the Coarse grid. 
A quantitative analysis is performed using direct indicators measuring the size of the recirculation region. $\Delta x_{r}/c$ is the maximum size in the streamwise direction, while $\Delta y_r / c$ indicates the maximum size in the the normal direction. These quantities are calculated observing the condition $\overline{u_x} = 0$ for the three simulations. Results are shown in \autoref{tab:recirc_features}. When compared with results from the Reference Grid, the Fine Grid simulation exhibits discrepancies of $8\,\%$ and $4\,\%$ for $\Delta x_r/c$ and $\Delta y_r/c$, respectively. Finally, the flow solution obtained using the Coarse Grid reaches discrepancies of $27\,\%$ and $7\,\%$. Lastly, $x_{\textrm{sep}}/c$ indicates the point on the airfoil where the flow is detaching from the body. Differences of $61\,\%$ are observed for the Coarse Grid, while the Fine Grid is reduced to $25\,\%$. In summary, one can see that numerical results exhibit a sensitivity to grid resolution. This feature will be exploited in the Data Assimilation techniques which will be used in the present work.

\begin{table}[h!]
    \centering
    \begin{tabular}{lcccc} \toprule[1.5pt]
         & $\alpha \, [\textrm{deg}]$ & $\Delta x_r/c$ & $\Delta y_r/c$  & $x_{\textrm{sep}}/c$\\ \cmidrule[0.25pt]{2-5} 
        Reference grid & $25$  &  $1.04$   & $0.079$    &  $0.165$  \\  
        Fine grid &  $25$  &      $0.96$   & $0.076$    & $0.22$  \\ 
        Coarse grid &  $25$  &    $0.76$   & $0.085$   &  $0.42$  \\ \bottomrule[1.5pt]  
    \end{tabular}
    \caption{Features of the recirculation region above the NACA\,0012 airfoil at an angle of attack $\alpha=25^\circ$ for the Coarse, Fine and Reference grid refinements.}
    \label{tab:recirc_features}
\end{table}

\section{Data-driven methodologies} \label{sec:data_driven_methodologies}

Data Assimilation (DA) \citep{Daley1991_cambridge,Asch2016_SIAM} is a branch of the Estimation Theory \cite{Simon2006_wiley} which aims to obtain accurate prediction of physical phenomena when multiple sources of measurement are available. These sources can provide limited information (in terms of variables of investigation), which can be scattered in space and / or time and affected by uncertainty. DA has been extensively used in meteorology and environmental sciences, but studies in fluid mechanics to applications of industrial interest have emerged in recent years \cite{Labahn2019_pci,ZHANG2021104962,Mons2021_jfm,Villanueva2024_IJHFF}. Classical applications, which are also relevant to the present work, combine two sources of information. The first one, referred to as \textit{model}, is a quasi-continuous representation of the physical process investigated in space and time. This prediction, which can for example the solution of a CFD solver in fluid mechanics, may be affected by strong uncertainties / biases. The seconds source, referred to as \textit{observation}, is usually a high-fidelity representation of the same phenomenon which is however limited in sampling in space and time. Experimental or high-fidelity simulation databases are a reliable source of observation. The way the prediction obtained by the different tools is combined depends on the DA strategy selected. Different approaches exhibit improved accuracy or reduced costs depending on the features of the physical process investigated.        

Variational DA, which includes well-known strategies such as 4D-Var and 3D-Var methods, resolves the DA problem as an optimization process via the minimization of a given cost function. These methods are very precise, but they usually require an attractor for the solution to exhibit robust convergence \cite{ide1997, rabier2000}. In application of fluid mechanics, they have mainly been used for the analysis of stationary or slowly evolving flow configurations \cite{baker2009}. 
Sequential methods are mainly based on Bayesian techniques \citep{Grudzien_Colin_Bocquet_Marc_2023}.  Among the approaches proposed in the literature, the Ensemble Kalman Filter (EnKF) \citep{evensen_sequential_1994} is a powerful tool to obtain state estimation for multiscale, unsteady physical processes. The EnKF is based on an ensemblistic formulation of the Kalman Filter (KF) \citep{kalman_new_1960}, which permits to approximate the evolution in time of an error covariance matrix. The latter is needed to successfully take into account uncertainty in the predictions provided by the model and the observation. Recent applications in fluid mechanics have shown its potential to the study of unsteady flows \cite{labahn2019, ZHANG2021104962}, such as the test case presented in \autoref{sec:numerical_tools}.

\subsection{Ensemble Kalman Filter}

The main features of the stochastic Ensemble Kalman Filters are now introduced. 
Let us consider the state vector $\mathbf{x}_{k-1}$ which includes physical information about the system studied produced by the model (such as the velocity and the pressure field in the elements of the grid) at an arbitrary simulation time step $k-1$. This state vector can be advanced in time from $k-1$ to $k$ using the model $\mathcal{M}$. Let us also consider that some observation $\mb{y}{}_{k}$ is available at the time step $k$. The EnKF produces a state estimation combining the two sources of information using a stochastic algorithm which takes into account the level of confidence in the available data.
The algorithm is organized into the following sequential operations:
\begin{enumerate}
    \item \textbf{Forecast step using the model.} $N_e$ model realizations $\state_{k-1}$ are advanced from time $k-1$ to $k$ using the model $\mathcal{M}$. This time advancement is referred to as the \textit{forecast} step ($f$ suffix). For the member $i$ of the ensemble realization, this corresponds to:
    \begin{equation}
    \mb{x}{f}_{i,k} = \model{}{\mb{x}{f}_{i,k-1}, \, \boldsymbol{\theta}_{i,k-1}} 
    \end{equation}
    $\boldsymbol{\theta}$ is an array including information of the parametric description of the model. In fluid mechanics this array can include model constants used in boundary conditions or turbulence closures.
\item \textbf{Manipulations in the observation's space.} 
$N_o$ scalar samples constitute the array $\mb{y}{}_{k}$ of observation at time $k$. This array is artificially perturbed $N_e$ times using a Gaussian noise,
\begin{equation}
    \mb{y}{}_{i,k} = \mb{y}{}_{k} + \bm{\vartheta}_i \, , \quad  i \in [1,N_e]  \quad ; \quad \bm{\vartheta}_i \sim \mathcal{N} \left( 0, \bm{\varsigma}_k \right) \quad ; \quad \bm{\varsigma_k} \in \mathbb{R}^{N_o \times N_o}
\end{equation}
where $\bm{\varsigma_k}$ quantifies the uncertainty in the observation. This matrix, which is very difficult to model without apriori information, is usually approximated to be diagonal in order to improve the robustness of the method. At last, the state $\mb{x}{f}_{i,k}$ predicted in the forecast step is projected into the observation space via the operator $\mathcal{H}$ at this stage. The term ${\obs{\mb{x}{}_{k}}}$ may represent an interpolation of the model data to the position of the sensors, for example if the observation has not been sampled in the center of a mesh element used by the model.
 \item \textbf{Calculation of the ensemble's mean.}
 The vectors ${\mb{x}{}}_{i,k}$ and ${\obs{\mb{x}{}_{k}}}$ are averaged over the ensemble members in the first step needed to approximate the error covariance matrix:
 \begin{equation}\label{eq:mean}
     \mean{\state^f}_k = \overline{\mathbf{x}}^f_k = \frac{1}{N_e}\sum_{i=1} ^{N_e} \mathbf{x}^f_{i,k} \,; \quad      \mean{\obs{\state{}^f}}_k = \overline{\obs{\mb{x}{f}_{k}}} = \frac{1}{N_e}\sum_{i=1} ^{N_e} \obs{\mb{x}{f}_{i,k}}
 \end{equation}
 \item \textbf{Calculation of the Covariance Matrices.} 
 First, Anomaly matrices are calculated. Each column of such matrix represents the deviation (or anomaly) of a single ensemble member from the ensemble mean. For the state array, one can write:
\begin{equation}\label{eq:anomaly}
\begin{aligned}
    \anomaly{\mathbf{x}^f}_k &= \frac{1}{\sqrt{N_e-1}}\left[ \mathbf{x}_0 ^f - \mean{\state^f}_k, \dots, \mathbf{x}_{N_e} ^f - \mean{\state^f}_k\right]\
    \end{aligned}
\end{equation}

Analogously \autoref{eq:anomaly} can be used to compute $\anomaly{\obs{\state^f}}_k$. The ensemble approximation of the Covariance Matrices is thus obtained:
\begin{equation}\label{eq:covariances}
    \begin{aligned}
    \bm{\Sigma}_k ^{\bm{x^f},\mathcal{H}(\mathbf{x^f})} = \bm{\Lambda}^{\mathbf{x}^f}\bm{\Lambda}^{\mathbf{\mathcal{H}\left(\mathbf{x}^f\right)}\intercal} \\
    \bm{\Sigma}_k ^{\mathcal{H}\left(\mathbf{x^f}\right),\mathcal{H}(\mathbf{x^f})} = \bm{\Lambda}^{\mathcal{H}(\mathbf{x^f})}\bm{\Lambda}^{\mathbf{\mathcal{H}\left(\mathbf{x}^f\right)}\intercal}
    \end{aligned}
\end{equation}
For brevity we will refer to the covariance matrices of two quantities $\bm{\varphi}$ and $\bm{\Gamma}$ as $\covariance{\bm{\varphi}}{\bm{\Gamma}}$ computed as in \autoref{eq:covariances}.
\item \textbf{Calculation of the Kalman Gain}.
The Kalman gain matrix $\mathbf{K}_k$ is the numerical ingredient permitting the combination of model results and observation. It takes into account the uncertainty in the data provided and it allows to obtain the DA state estimation. It is obtained as a manipulation of the covariance matrices:
\begin{equation}\label{eq:kalman_gain}
    \mathbf{K}_k =  \bm{\Sigma}_k ^{\bm{x^f},\mathcal{H}(\mathbf{x^f})} \left[  \bm{\Sigma}_k ^{\mathcal{H}\left(\mathbf{x^f}\right),\mathcal{H}\left(\mathbf{x^f}\right)} + \bm{\varsigma}_k \right]^{-1}
\end{equation}
\item \textbf{Analysis phase.} The state estimation is here obtained via an algebraic operation which combines the arrays and matrices previously generated. The resulting state prediction (analysis, suffix $a$) for each ensemble member $i$ is obtained as: 
 \begin{equation}
    \mathbf{x}_{i,k}^a = \mathbf{x}_{i,k}^f + \mathbf{K}_k \left( \mathbf{y}_{i,k} - \mathcal{H}(\mathbf{x}_{i,k}^f) \right) 
    \label{eq:analysis}
\end{equation}

\end{enumerate}

The procedure is then repeated performing multiple forecast steps until observation is available at a specific time step. Whenever optimization of the model parameters $\boldsymbol{\theta}$ is required, this is performed using the \textit{extended state} procedure \cite{Asch2016_SIAM}. The EnKF procedure is here performed for a new state $\mathbf{x^\prime} =  [\mathbf{x} \, \boldsymbol{\theta}]^\top$ with no modification to the DA equations. 
The main advantages of this method are that:
\begin{enumerate}
    \item it is efficient with non-linear physical processes thank to the independent forecast steps.
    \item it excludes prohibitively expensive time advancement and inversion of the complete covariance matrix used in the classical KF approaches. 
\end{enumerate}  
However, if the state array is large, computations in \autoref{eq:analysis} can be nonetheless extremely expensive. In the following sections, procedures to speed up calculations while preserving a satisfactory level of accuracy are presented.


\subsection{Multi-grid Ensemble Kalman Filter}
\label{sec::MGEnKF}
Despite the favorable features of the EnKF for the analysis of complex dynamic systems using numerical tools, CFD applications require the use of a number of grid elements which can become prohibitively expensive in computational terms in particular as ensemble realizations are required. The Multi-grid Ensemble Kalman Filter (here indicated MGEnKF) is a methodology based on the EnKF which exploits multi-level calculation performed in CFD solvers. Initially developed by \citet{moldovan_multigridensemble_2021}, the MGEnKF relies on two model predictions of the state $\mathbf{x}$. The first one, which is the final output of the CFD model $\mathcal{M}^\textbf{F}$, is calculated on the relatively well refined grid (\textit{Fine grid}). The second one is the result of calculations on a \textit{Coarse grid}, which corresponds to a reduced version of the \textit{Fine grid} and it is commonly used for initial iterations. The model resolved on this Coarse grid is $\mathcal{M}^\textbf{C}+ \mathcal{C}$. The term $\mathcal{M}^\textbf{C}$ corresponds to the dynamic model resolved on the Coarse grid, while the term $\mathcal{C}$ is a correction term which can be included to compensate the loss in accuracy due to grid coarsening. This technique generates ensemble members on the coarser grid used for the model, which are used for DA purposes, and it informs a single model run performed on the \textit{Fine grid}. Therefore, DA costs for field manipulations and memory storage are dramatically reduced. The MGEnKF algorithm involves two distinct procedures on the Coarse grid. The first one, referred to as \textit{outer loop}, performs the DA procedure using the ensemble members on the Coarse grid and available observation. The second one, referred to as \textit{inner loop}, samples data from the single solution on the Fine grid and uses it as surrogate observation for a second DA cycle using the ensemble members \cite{moldovan_optimized_2022}. The inner loop usually targets the optimization of the corrective term $\mathcal{C}$ manipulating some available free coefficients $\boldsymbol{\psi}$, so that the discrepancy between the accuracy of the prediction using Coarse and Fine grid is reduced. Using the same parameters previously introduced for the EnKF, the MGEnKF algorithm is structured in the following operations:


\begin{enumerate}
    \item \textbf{Forecast.} The $N_e$ ensemble members calculated on the Coarse grid ($\mathbf{C}$)  are advanced in time using the Coarse-grid model $\mathcal{M}^\textbf{C}$:
    \begin{equation}
        \mathbf{x^C}_{i,k}^f = \mathcal{M}^\textbf{C} \left(\mathbf{x^C}_{i,k-1}^a, \, \boldsymbol{\theta}_{i,k-1}\right) + \mathcal{C} \left(\mathbf{x^C}_{i,k-1}^a, \, \boldsymbol{\psi}_{i,k-1}\right) \label{eq:mgenkf_forecast}
    \end{equation}
    
    Similarly, the fine-grid ($\mathbf{F}$) solution $\mathbf{x^F}$ is advanced in time thanks to the corresponding fine-grid model $\mathcal{M}^\textbf{F}$:
    \begin{equation}
        \mathbf{x^F}_{k}^f = \mathcal{M}^\textbf{F} \left(\mathbf{x^F}_{k-1}^a\right)
    \end{equation}

    \item \textbf{Projection from Fine-grid to Coarse-grid.} The forecast previously obtained on the Fine grid is projected in the Coarse-grid space using the projection operator $\bm{\Phi}^\star_r$:
    \begin{equation}
    \widetilde{\mathbf{x^C}}_{k}^f = \bm{\Phi}^\star_r \left(  \mathbf{x^F}_{k}^f  \right) \label{eq:mgenkf_proj}
    \end{equation}

    \item \textbf{Inner loop: DA optimization of the coarse model.} The free coefficients $\boldsymbol{\psi}_{i,k-1}$ are optimized using surrogate observation sampled from the field $\widetilde{\mathbf{x^C}}_{k}^f$.

    \item \textbf{Outer loop: EnKF calculations on the Coarse grid.} 
    The ensemble members are updated using available external observation $(\mathbf{y}_{i,k}$ via the Kalman gain $\mathbf{K^C}$:
    \begin{equation}
        \mathbf{x^C}_{i,k} ^a = \mathbf{x^C}^f_{i,k} + \mathbf{K^C}\left(\mathbf{y}_{i,k} - \mathcal{H}\left(\mathbf{x^C}_{i,k} ^f\right)\right) \label{eq:mgenkf_outer_loop}
    \end{equation}
    The Kalman gain is also used to update the projected state:
     $\widetilde{\mathbf{x^C}}_k ^f$:
        \begin{equation}
        \widetilde{\mathbf{x^C}}_{k} ^a = \widetilde{\mathbf{x^C}}^f_k + \mathbf{K^C}\left(\mathbf{y}_k - \mathcal{H}\left(\widetilde{\mathbf{x^C}}_k ^f\right)\right) \label{eq:mgenkf_kalman_gain}
    \end{equation}

    \item \textbf{Update of the fine-grid state.} The Coarse grid state updates are projected back to the full order space with the projection operator $\bm{\Phi}_r$ to update the fine-grid solution:
    \begin{equation}
        \mathbf{x^F}_{k}^a = \mathbf{x^F}^f_k + \bm{\Phi}_r \left( \widetilde{\mathbf{x^C}}_k^a - \widetilde{\mathbf{x^C}}^f_k\right) \label{eq:mgenkf_fine_grid_update}
    \end{equation}
\end{enumerate}

\begin{figure}[h!]
    \centering
    \includegraphics[width=0.75\linewidth]{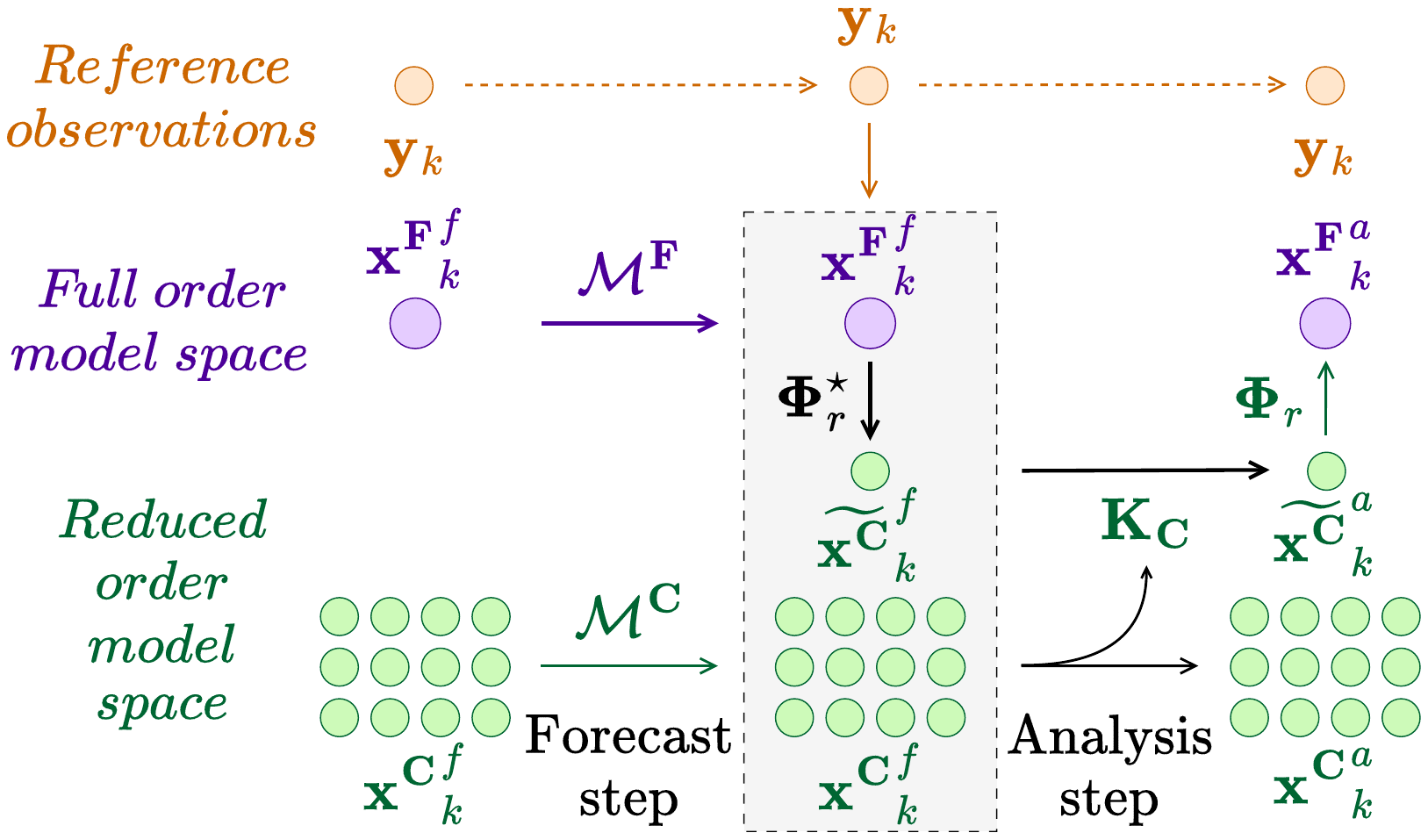}
    \caption{MGEnKF schematic representation of the outer loop.}
    \label{fig:MGEnKF_scheme}
\end{figure}

A qualitative representation of the MGEnKF algorithm is shown in \autoref{fig:MGEnKF_scheme}.

\subsection{Multi-fidelity Ensemble Kalman Filter with Reduced Order Control Variates}


\citet{popov_multifidelity_2021} recently proposed a dedicated DA framework known as Multi-Fidelity Ensemble Kalman Filter, which uses variance reduction strategies for Monte-Carlo method. This method combines a full-order physical model (FOM) with a hierarchy of reduced-order surrogate models (ROM) to increase the computational efficiency of Data Assimilation. A key aspect of this approach is its foundation in the theory of linear control variates, a variance reduction technique commonly used in Monte-Carlo methods \citep{rubinstein_efficiency_1985}.

Let us consider a high-fidelity model, whose state is represented by $\bm{\chi}$ and a lower-fidelity model, with state $\bm{\gamma}$.  $\bm{\gamma}$ is said to be a control variate of $\bm{\chi}$ if $\bm{\gamma}$ exhibits a strong correlation with $\bm{\chi}$ and its expectation $\bm{\mu}_{\bm{\gamma}}$ is known. The primary goal of control variates is to estimate the statistical properties of the high fidelity state including information derived from the lower-fidelity state, which is computationally less expensive. 
To this purpose, the control variate algorithm uses a total variate $\bm{\xi}$ defined as a linear combination of the ensemble means $\bm{\mu}_{\bm{\chi}}$ and $\bm{\mu}_{\bm{\gamma}}$ of $\bm{\chi}$ and $\bm{\gamma}$, respectively. 
Given these conditions, $\bm{\xi}$ is:
\begin{equation}
    \bm{\xi} = \bm{\chi} - \lambda \left( \bm{\gamma} - \bm{\mu}_{\bm{\gamma}} \right) \label{eq:total_variate}
\end{equation}

The random variable $\bm{\chi}$ is referred to as the \textsl{principal} variate, $\bm{\gamma}$ is the \textsl{control} variate and finally $\bm{\xi}$ is the \textsl{total} variate \citep{popov_multifidelity_2021}. In \autoref{eq:total_variate}, the variable $\bm{\xi}$ has a reduced variance compared to $\bm{\chi}$, whereas its mean is equal to the principal variate mean $\bm{\mu}_{\bm{\xi}} = \bm{\mu}_{\bm{\chi}}$. Finally, $\lambda$ is chosen here as $\lambda \in \mathbb{R}$ so to minimize the variance of the total variate.  In most of practical cases $\bm{\mu}_{\bm{\gamma}}$ is unknown, therefore a third variable $\bm{\eta}$ (referred to as the \textsl{ancillary} variate) is introduced along the principal and control variates. This variable, for which ensemble calculations can be performed, is selected so that $\bm{\mu}_{ \bm{\gamma}} = \bm{\mu}_{ \bm{\eta}}$. This usually implies that the same model is used to obtain $\bm{\gamma}$ and $\bm{\eta}$. Using this approximation  \autoref{eq:total_variate} becomes:
\begin{equation}
        \bm{\xi} = \bm{\chi} - \lambda \left( \bm{\gamma} - \bm{\eta} \right) \label{eq:total_variate_2}
\end{equation}
Let now transpose \autoref{eq:total_variate_2} to a multi-level system with model runs using a Fine grid - Coarse grid such as the one used for the MGEnKF. The algorithm is extremely similar for the two methods, but three sub-ensembles must be defined. The first one corresponds to the principal variate $\bm{\chi}$, for which $N_{\bm{\chi}}$ members are run. 
These simulations are performed on the Fine cascade of the system. For the control variate $\bm{\gamma}$, $ N_{\bm{\gamma}} = N_{\bm{\chi}}$ ensemble members are generated. It will be shown in the following that this choice is necessary to preserve correlation for the representations of $\bm{\chi}$ and $\bm{\gamma}$. However, these simulations are run on the Coarse grid. Finally, $N_{\bm{\eta}}$ ensemble members are run on the Coarse grid for the ancillary variate $\bm{\eta}$. One can see that $N_{\bm{\chi}}$ runs are performed on the Fine grid and $N_{\bm{\gamma}}+N_{\bm{\eta}}$ on the Coarse grid. Therefore, the total number of ensemble members $N_e$ is equal to:
\begin{equation}
    N_e = 2 N_{\bm{\chi}} + N_{\bm{\eta}}
\end{equation}

\begin{figure}[h!]
    \centering
    \includegraphics[width=0.75\linewidth]{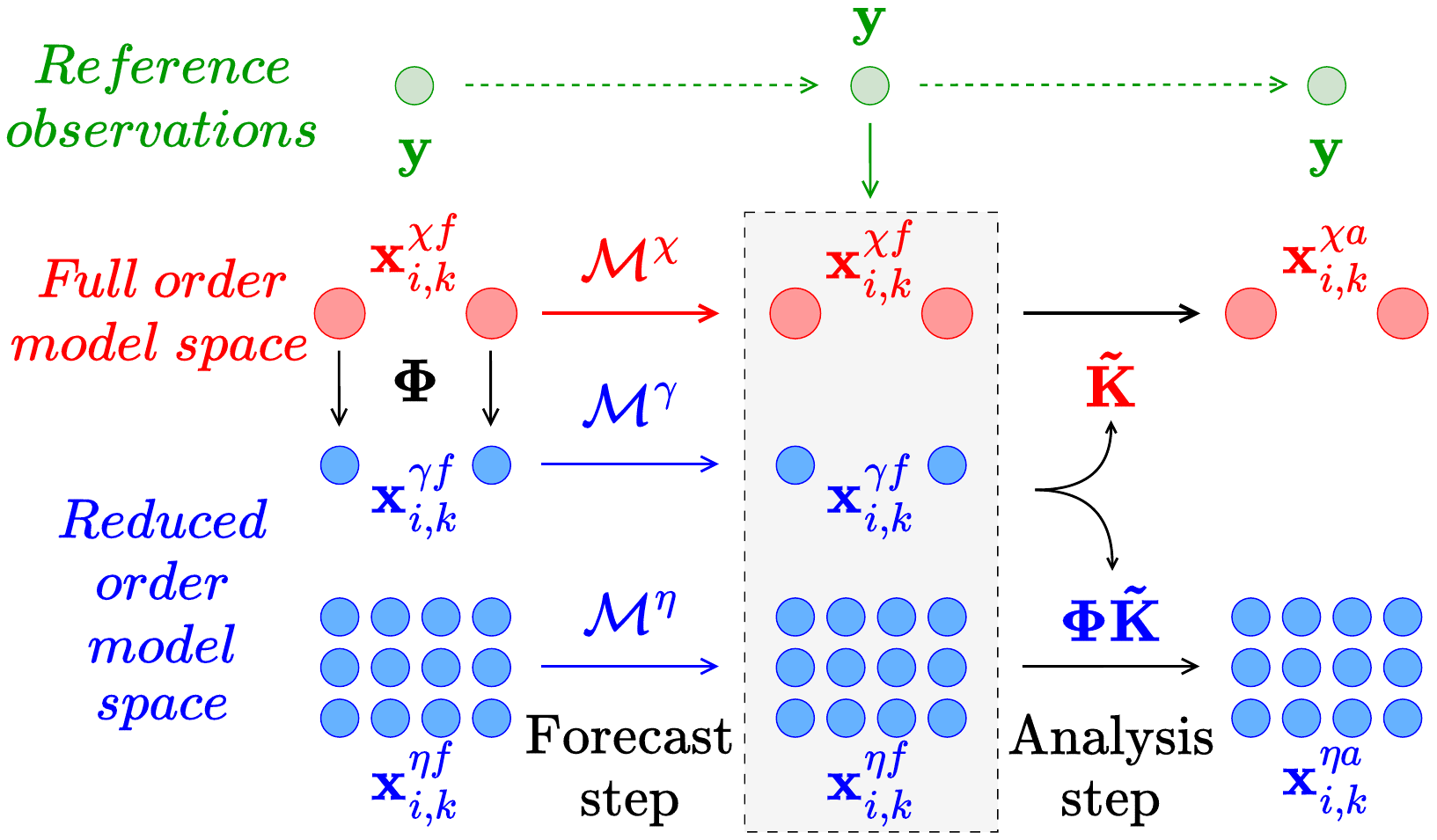}
    \caption{MFEnKF schematic representation.}
    \label{fig:MFENKF_scheme}
\end{figure}

This implies that state estimation can be obtained with a limited amount of $ N_{\bm{\chi}}$ ensemble members performed on the Fine grid and a significantly larger number of runs on the Coarse grid equal to $ N_{\bm{\gamma}}+ N_{\bm{\eta}}$. The steps of MFEnKF algorithm are now discussed and a qualitative representation is provided in \autoref{fig:MFENKF_scheme}. 

\begin{enumerate}
    \item \textbf{Forecast step.} Members from the three ensemble are advanced in time according to their corresponding non-linear models $\mathcal{M}$:

    \begin{equation}\label{eq:mfenkf_forecast}
    \begin{aligned}
        \principal{f}_{i,k}&= \model{\bm{\chi}}{\principal{a}_{i,k-1}} \\
        \control{f}_{i,k} &= \model{\bm{\gamma}}{\control{a}_{i,k-1}} \\
        \ancillary{f}_{i,k} &= \model{\bm{\eta}}{\ancillary{a}_{i,k-1}} \\
    \end{aligned}
    \end{equation}
    \item \textbf{Projection - $\bm{\chi}$ to $\bm{\gamma}$.} Due to the differences between $\mathcal{M}^{\bm{\chi}}$ and $\mathcal{M}^{\bm{\gamma}}$, owing to different grid resolution in the present study, physical fields obtained with principal and control members will decorrelate.  
    Solutions from the principal variate ensemble are projected to the control variates using the projection operator $\bm{\Phi}^{\star}$ that maps the Fine grid solutions to the Coarse grid:
    \begin{equation}
        \control{f}_{i,k} = \bm{\Phi^*}\left(\principal{f}_{i,k}\right)
    \end{equation}
    \item \textbf{Total variate mean.} In that current ensemblist framework, the ensemble average of the three forecast ensemble variates $\mean{\principal{f}}, \mean{\control{f}},\mean{\ancillary{f}}$ is used to compute the total variate mean estimation: 
    \begin{equation}
    \begin{aligned}
    \mean{\total{f}}_k &= \mean{\principal{f}}_k - \lambda \left( \mean{\control{f}}_k - \mean{\ancillary{f}}_k\right)\\
    \end{aligned}
    \end{equation}

    \item \textbf{Total variate observation operator.} We use the full-space observation operator $\mathcal{H}$ for the principal variate and the same is applicable for control and ancillary variates projected via the operator $\bm{\Phi}$. However the total variate $\total{f}$ is not observed directly. To overcome this limitation an operator $\overline{\mathcal{H}}$ with the following formulation is suggested:  
       \begin{equation}
            \begin{aligned}
            \overline{\obs{\total{f}_{i,k}}} = \obs{\principal{f}_{i,k}} - \frac{1}{2}\obs{\bm{\Phi}\left(\control{f}_{i,k}\right)} + \frac{1}{2}\obs{\bm{\Phi}\left(\ancillary{f}_{i,k}\right)}
            \end{aligned}
        \end{equation}
    \item \textbf{Calculation of the covariance matrices.} Anomaly matrices are used to construct the approximated covariance matrices of the total variate assuming principal and ancillary as random and independent variables:
    \begin{equation}
        \begin{aligned}
        \covariance{\total{f}}{\overline{\obs{\total{f}}}} &= \covariance{\principal{f}}{\obs{\principal{f}}}\\&+ \frac{1}{4}   \left(\bm{\Phi}\left(\covariance{\control{f}}{\obs{\bm{\Phi}\left(\control{f}\right)}}\right) + \bm{\Phi}\left( \covariance{\ancillary{f}}{\obs{\bm{\Phi}\left(\ancillary{f}\right)}}\right)\right) \\&-\frac{1}{2}\left(\covariance{\principal{f}}{\obs{\bm{\Phi}\left(\control{f}\right)}}
        +\bm{\Phi}\left(\covariance{\control{f}}{\obs{\principal{f}}}\right)\right) \\
        \covariance{\overline{\obs{\total{f}}}}{\overline{\obs{\total{f}}}} &= \covariance{\obs{\principal{f}}}{\obs{\principal{f}}}\\
        &+ \frac{1}{4}\left( \covariance{\obs{\bm{\Phi}\left(\control{f}\right)}}{\obs{\bm{\Phi}\left(\control{f}\right)}}+\covariance{\obs{\bm{\Phi}\left(\ancillary{f}\right)}}{\obs{\bm{\Phi}\left(\ancillary{f}\right)}}\right)\\
        &-\frac{1}{2}\left(\covariance{\obs{\principal{f}}}{\obs{\bm{\Phi}\left(\control{f}\right)}}+\covariance{\obs{\bm{\Phi}\left(\control{f}\right)}}{\obs{\principal{f}}}\right) 
        \end{aligned}
    \end{equation}
    Finally the observation error covariance matrices are built using independent random variables so that:
    \begin{align}
        \nonumber \bm{\theta}^{\bm{\chi}} &\sim \mathcal{N}\left(0,\bm{\varsigma}^{\bm{\chi}}\right)\\
        \bm{\theta}^{\bm{\gamma}} &= \bm{\theta}^{\bm{\chi}}\\
        \nonumber \bm{\theta}^{\bm{\eta}} &\sim \mathcal{N}\left(0,\bm{\varsigma}^{\bm{\eta}}\right)
    \end{align}
    According to Popov et al. \cite{popov_multifidelity_2021} this leads to:
    \begin{equation}
        \covariance{\bm{\theta}^{\bm{\chi}}}{\bm{\theta}^{\bm{\chi}}} =         \covariance{\bm{\theta}^{\bm{\gamma}}}{\bm{\theta}^{\bm{\gamma}}} =        \covariance{\bm{\theta}^{\bm{\eta}}}{\bm{\theta}^{\bm{\eta}}} = 2        \covariance{\bm{\theta}^{\bm{\xi}}}{\bm{\theta}^{\bm{\xi}}}
    \end{equation}
    The perturbed observations $\textbf{y}^{\bm{\chi}}, \textbf{y}^{\bm{\eta}}, \textbf{y}^{\bm{\gamma}}$ are obtained using the corresponding random variables.
    
    \item \textbf{Kalman Gain Calculation.} The Kalman Gain is then computed to estimate the assimilated (a) total variate:
    \begin{equation}
    \begin{aligned}
        \tilde{\mathbf{K}} &= \covariance{\total{b}}{\overline{\obs{\total{b}}}}\left(\covariance{\obs{\total{b}}}{\obs{\total{b}}} + \covariance{\theta^{\bm{\xi}}}{\theta^{\bm{\xi}}}\right)^{-1}\\
        \total{a} &= \total{b} + \tilde{\mathbf{K}}\left(\overline{\obs{\total{b}}}- \mathbf{y}^{\bm{\xi}} \right)
    \end{aligned}
    \end{equation}
        \item \textbf{Variate ensemble update.}
        The MFEnKF assimilate the ensembles as follows:
        \begin{equation}
            \begin{aligned}
            \principal{a}_{i,k} &= \principal{f}_{i,k} + \tilde{\mathbf{K}}\left(\mathbf{y}^{\bm{\chi}}_{i,k}-\obs{\principal{f}_{i,k}}\right)\\
            \control{a}_{i,k} &= \control{f}_{i,k} + \bm{\Phi}\left(\tilde{\mathbf{K}}\left(\mathbf{y}^{\bm{\gamma}}_{i,k}-\obs{\control{f}_{i,k}}\right)\right)\\
            \ancillary{a}_{i,k} &= \ancillary{f}_{i,k} + \bm{\Phi}\left(\tilde{\mathbf{K}}\left(\mathbf{y}^{\bm{\eta}}_{i,k}-\obs{\ancillary{f}_{i,k}}\right)\right)\\
            \end{aligned}
        \end{equation}

\end{enumerate}

The control variate algorithm applied to an ensemblist framework enables the total variate $\total{}$ to have the same mean as the principal variate $\principal{}$, having smaller covariance, indicating therefore a higher confidence in the total variate results. 

\subsection{Convolutional Neural Networks}

It was shown in \autoref{sec::MGEnKF} that the inner loop of the MGEnKF can be used to optimize a subgrid correction term $\mathcal{C}$, governed by a set of coefficients $\boldsymbol{\psi}$, to be integrated in the simulations performed on the Coarse grid. \citet{moldovan_optimized_2022} investigated the DA optimization of the coefficients $\boldsymbol{\psi}$, showing that improvement of the accuracy for the model runs on the Coarse grid can be obtained. However, these functional forms are usually difficult to be deduced for complex flows. In the present work, $\mathcal{C}$ is obtained training complete data-driven models, therefore bypassing the need to provide a functional form. The data-driven strategy investigated relies on the use of Machine Learning (ML) and more precisely of a well-known family of Neural Networks. Mathematical frameworks to perform rigorous integration of DA and ML approaches have been recently proposed in the literature with the objective to emulate a dynamic sparse model \cite{brajard_combining_2020}, to infer unresolved scales \cite{brajard_combining_2021} or to improve turbulence modelling \citep{Villiers2025_ftc}.  

The ML tools used in this study are now presented. Convolutional Neural Networks, often referred to as CNNs or ConvNets, are a family of Machine Learning methods mainly used in the domain of computer vision \cite{lecun_deep_2015}. First conceptualized and successfully assessed by \citet{lecun_gradient_based_1998}, CNNs have rapidly evolved in the past 25 years with applications ranging from various face and object recognition \cite{garcia_convolutional_2004, ciresan_multi-column_2012} to medical imaging \cite{abd-alhalem_cervical_2024, elkholy_deep_2024}, as well as geophysical research for earthquakes \cite{rouet-leduc_autonomous_2021}. Fluid mechanics studies have also seen application of CNNs, both for experimental \cite{zhang_pyramidal_2023} and numerical investigations \cite{lecler_prediction_2023, fukami_super-resolution_2023, rana_scalable_2024}. 

\begin{figure}[h!]
    \centering
    \includegraphics[width=0.99\linewidth]{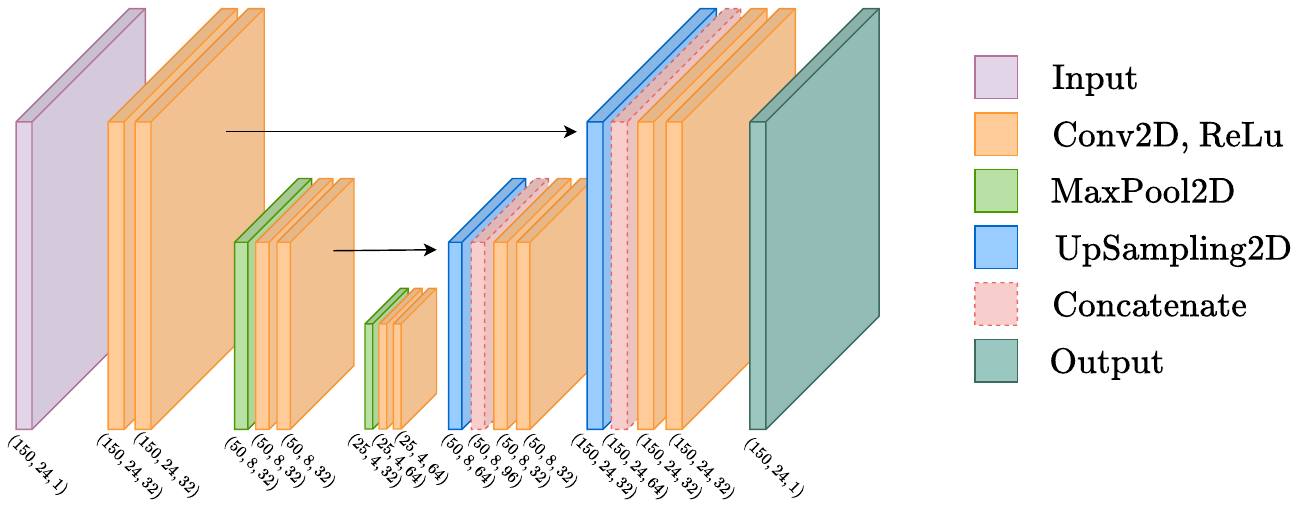}
    \caption{CNN architecture and its corresponding channel dimensions.}
    \label{fig:CNN_struct}
\end{figure}

When dealing with CNNs, several architectures are available. A complete description of the main architectures and their associated techniques is reviewed in \citet{zhao_review_2024}. Within the scope of this study, the Machine Learning inference will act as a regressor and the CNN architecture must be selected accordingly. \citet{ronneberger_u-net_2015} developed a fully-convolutional architecture which is frequently named \textit{U-net} because of its visual representation when sketched. This architecture is characterized by successive reductions in the spatial dimensions (downsampling) and data expansions (upsampling) operations. The downsampling process, or \textbf{encoder}, extracts information at different scales, reducing spatial dimensions while increasing features depth. The upsampling process, or \textbf{decoder}, progressively recovers the spatial dimensions while enhancing the spatial localization of the flow features. The main structure of the CNN is sketched in \autoref{fig:CNN_struct}. 

As it has been explained, \textit{U-nets} are built up from a contraction, followed by an expansion process. On the one side, the contraction consists in convolution layers, activated by a Rectified Linear Unit (ReLU) function before being down-sampled thanks to a maximum pooling operation. In the present study, this procedure is carried out twice. Then, on the other side, the expansion process is performed with up-sampling layers, followed by concatenation layers before applying new convolutions which are also activated with the ReLU function. As two contraction operations are performed during the encoding process, it is mandatory to carry out the same number of upsampling operations during decoding. The final layer is a convolution layer with a channel depth equal to one, with the objective of obtaining an output whose dimensions are the same as the input. The black left-to-right arrows specify from which location of the \textit{U-net} the concatenations are executed. From the channel dimensions given on \autoref{fig:CNN_struct} and \autoref{tab:cnn_tab_description}, this CNN \textit{U-net} architecture is resulting in a number of $592\,513$ trainable parameters. Finally, this CNN architecture was built with the TensorFlow Python library \cite{tensorflow2015-whitepaper} using the Keras backend \cite{chollet2015keras}. 

\begin{table}[h!]
\scriptsize
\centering
\begin{booktabs}{lcc}
\toprule[1.5pt]
\textbf{Layers}    & \textbf{Output shape} & \textbf{Number of parameters} \\ \midrule[0.25pt]
Input     &   $(150,24,1)$  &    $0$        \\ \midrule[dashed]
Conv2D    &   $(150,24,32)$ &    $1\,184$   \\
ReLU      &   $(150,24,32)$  &  $0$   \\
Conv2D    &   $(150,24,32)$  &   $36\,896$   \\
ReLU      &   $(150,24,32)$  &   $0$  \\
MaxPool2D &   $(50,8,32)$  &      $0$  \\ \midrule[dashed]
Conv2D    &   $(50,8,32)$  &    $36\,896$      \\
ReLU      &   $(50,8,32)$   &   $0$     \\
Conv2D    &   $(50,8,32)$  &    $36\,896$  \\
ReLU      &   $(50,8,32)$  &  $0$   \\
MaxPool2D &   $(25,4,32)$  &  $0$    \\ \midrule[dashed]
Conv2D    &   $(25,4,64)$   & $73\,792$  \\
ReLU      &   $(25,4,64)$   &  $0$  \\
Conv2D    &   $(25,4,64)$  &  $147\,520$ \\
ReLU      &   $(25,4,64)$   & $0$  \\ \midrule[dashed]
UpSampling2D & $(50,8,64)$  & $0$ \\
Concatenate  & $(50,8,96)$  & $0$ \\
Conv2D    &  $(50,8,32)$ &  $110\,624$    \\
ReLU    &   $(50,8,32)$  & $0$  \\
Conv2D  &    $(50,8,32)$ & $36\,896$  \\
ReLU        &  $(50,8,32)$  &  $0$  \\  \midrule[dashed]       
UpSampling2D & $(150,24,32)$  & $0$ \\
Concatenate  & $(150,24,64)$  & $0$ \\        
Conv2D    &  $(150,24,32)$ &  $73\,760$    \\
ReLU    &   $(150,24,32)$  & $0$  \\
Conv2D  &    $(150,24,32)$ & $36\,896$  \\
ReLU        &  $(150,24,32)$  &  $0$          \\   \midrule[dashed]
Conv2D (output)  &    $(150,24,1)$ & $1\,153$  \\   
\bottomrule[1.5pt]      
\end{booktabs}
\caption{CNN \textit{U-net} structural description}
\label{tab:cnn_tab_description}
\end{table}

\section{Assessment of multi-level and multi-fidelity EnKF strategies}\label{sec:data_driven_optimization_inlet_BC}

The EnKF models presented in \autoref{sec:data_driven_methodologies} are used to investigate the flow around a cascade of NACA\,0012 profiles. Four DA runs will be performed and results will be compared in terms of accuracy and computational resources required. 
\subsection{Set up used for the EnKF}

\begin{figure}[h!]
    \centering
    \includegraphics[width=0.75\textwidth]{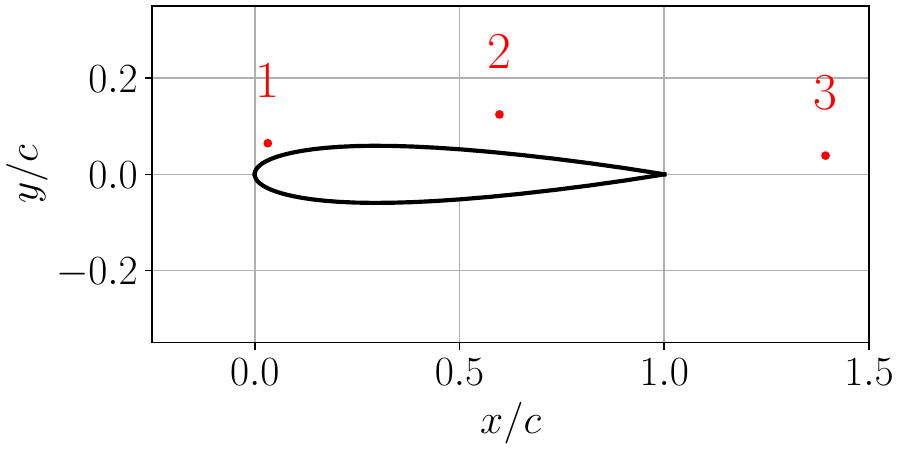}
    \caption{Location of the sensors around the NACA\,0012 profile.}
    \label{fig:probes_location}
\end{figure}

The key elements used for the DA investigations are now presented:
\begin{itemize}
\item \textbf{Observation} is obtained sampling the instantaneous velocity field of a numerical simulation which is run using the Reference grid. This simulation, which will be referred to as ground truth, has been run selecting an angle of attack $\alpha=25^\circ$. Three sensors are used to sample the streamwise velocity $u_x$, the normal velocity $u_y$ and the pressure $p$ as shown in \autoref{fig:probes_location}. The coordinates of the sensors are $n_1 =[0.032, 0.064]$, $n_2=[0.6,0.12]$ and $n_3=[1.39,0.039]$ in $c$ units. The first sensor is located near the NACA\,0012 leading edge. At this location the boundary layer is attached. The second sensor is located further along the profile, and it can observe flow recirculation or not depending on the value of $\alpha$. The third sensor is located in the wake region. The observation vector consists of six scalar quantities:
\begin{equation}
    \mathbf{y} = \begin{pmatrix}
    u_{x,n_1}, u_{x,n_2}, u_{x,n_3},  u_{y,n_1}, u_{y,n_2}, u_{y,n_3}
    \end{pmatrix}
\end{equation} 
Uncertainty in the observation is obtained perturbing the measurements with an additive Gaussian noise 
$\mathcal{N} \left( \mathbf{y}, 4 \right)$
so that the noise is the same for every sensor and the observation's  normalized by $a_\infty$ variance is $0.012$. Data is sampled every $\Delta_{DA} = 13.6\, t_A$ which corresponds to $800$ time steps of the reference simulation. 
\item \textbf{The model} is the OpenFOAM native unsteady compressible solver \texttt{rhoPimpleFoam}, which is run using either the Fine grid or the Coarse grid. The state $\state{}$ is a vector of size $2\times n$ where $n$ is the number of cells of the discretized domain, which is therefore different when using the Fine grid or the Coarse grid. The state vector contains the two velocity components ($u_x$ and $u_y$)
as following:
\begin{equation}
\state{} = \begin{pmatrix}
u_{x,0},
\dots
u_{x,i},
\dots
u_{x,n}.
\dots
u_{y,0}
\dots
u_{y,i},
\dots
u_{y,n}
\end{pmatrix}
\end{equation}
\end{itemize}

\begin{figure}[h!]
    \centering
    \includegraphics[width=0.95\textwidth]{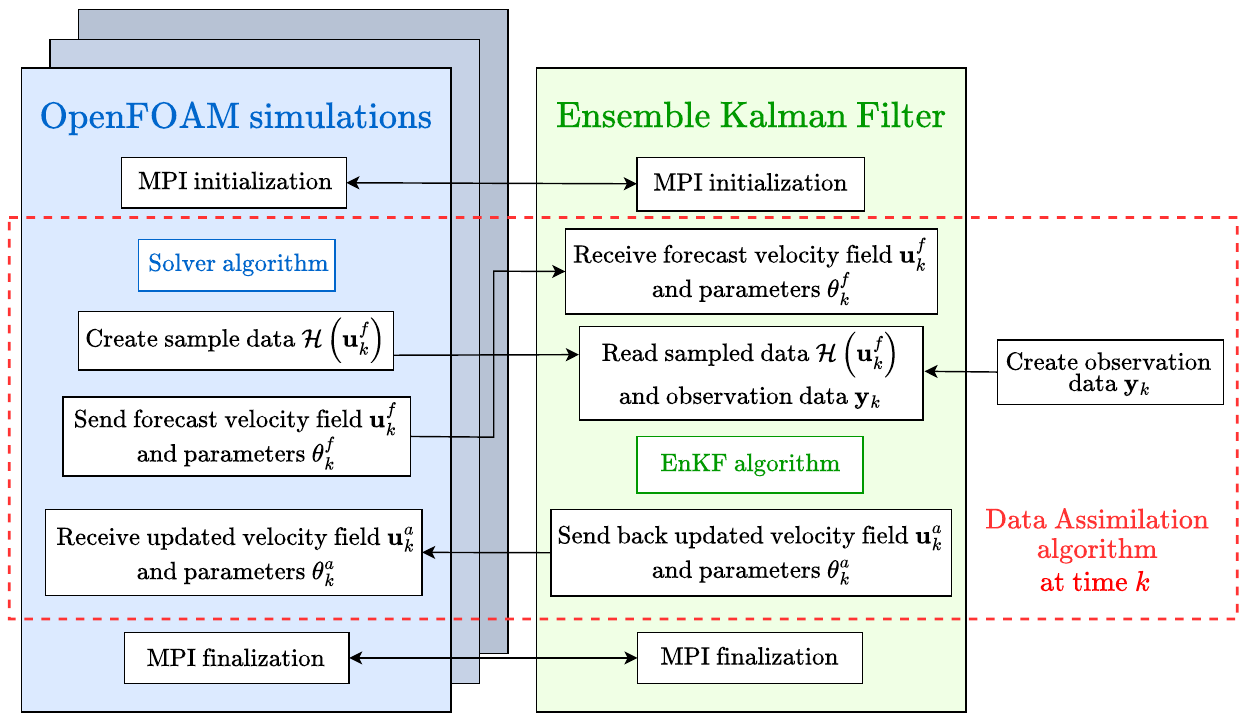}
    \caption{Schematic representation of CONES for online sequential Data Assimilation.}
    \label{fig:CONES_scheme}
\end{figure}

The Data Assimilation algorithm is now presented. Ensemble methods rely on the generation of several members which can rapidly become computationally expensive. In addition, carrying out a large number of EnKF analysis phases via stop-and-restart of the solver is not a long-term viable solution in terms of computational efficiency and human intervention. This study performs the DA tasks using the in-house developed framework CONES (\textbf{C}oupling \textbf{O}penFOAM with \textbf{N}umerical \textbf{E}nvironment\textbf{S}) which couples the CFD solvers of OpenFOAM with an EnKF numerical code to perform \textit{online} DA analysis \cite{villanueva_augmented_2023, valero_immersed_2025}. CONES parallelizes the DA operations via MPI communication. The library allocates CPU cores to OpenFOAM, automatically taking into account the number of simulations and the domain decomposition selected  by the user. Multiple additional CPU core is reserved to execute Data Assimilation algorithms. 
This architecture allows overall time-efficiency by eliminating the human operations and reducing machine queuing time required for manual simulation restart between each analysis phase. Moreover, by automating the Data Assimilation process, it removes the ROM memory overhead associated with storing data for \textit{offline} DA analysis, shifting it to RAM. Details of the library are given in \autoref{fig:CONES_scheme} and the source code is openly available at \url{https://gitlab.ensam.eu/pe431/cones-dev}. 

The DA runs using different versions of the EnKF will be used to optimize features of the inlet boundary condition. More specifically, the angle of attack $\alpha$, whose value has been fixed to $\alpha=25^\circ$ for the reference simulation, is supposed to be unknown and it will be inferred using the DA techniques. It is here reminded that this optimization targets to reduce the discrepancy between model prediction and observation, accounting for the prescribed uncertainty in the two sources of information. The period between two successive analysis steps $\Delta_{DA} = 13.6 \, t_A$ has been selected to permit the advection of the updated inlet features to the sensors \cite{moussie_statistical_2024}. During this period, the flow advection transports the optimized information at the inlet (in terms of updated $\alpha$ values) over a length of $\approx 13.6 \, c$, which is more than $50\,\%$ of the length of the physical domain investigated. In order to obtain a satisfying optimization of the coefficient $\alpha$, it is important to propagate modifications across the domain of the simulation and to dissipate unstationary effects associated with the change of the inlet conditions. The latter are in particular observed during the very first analysis steps, where some ensemble members are run with $\alpha$ values which can be far from the targeted value. At last, localization and inflation procedures \cite{Asch2016_SIAM} have not been used for the DA analyses performed in this work. Preliminary analyses for this test case exhibited a very small level of sensitivity of the solution to these techniques, and they are therefore excluded for sake of simplicity. 


\subsection{Sensitivity analysis using a single grid for the DA model}
\begin{figure}[h!]
    \centering
    \begin{tabular}{c}
    \includegraphics[width=0.65\linewidth]{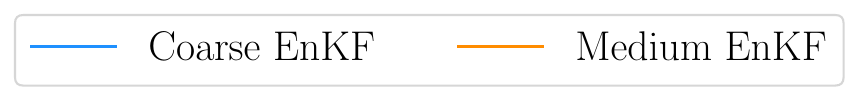} \\
    \includegraphics[width=0.65\linewidth]{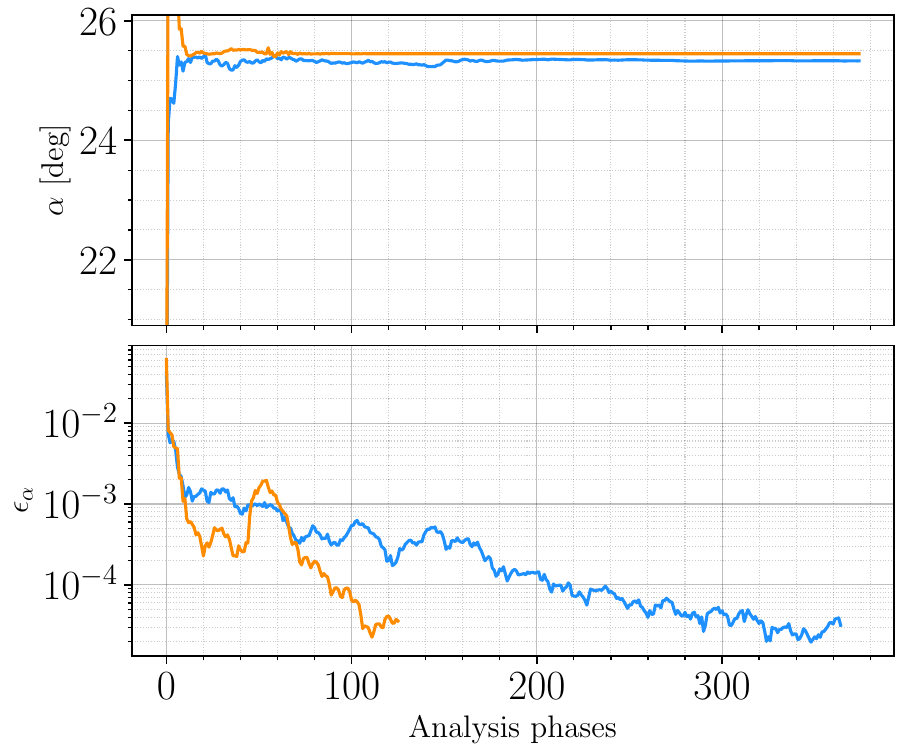}
    \end{tabular}
    \caption{Top: Evolution of the ensemble average angle of attack $\alpha$ applied to the inlet boundary condition. Bottom: Evolution of the convergence criterion $\epsilon_\alpha$.}
    \label{fig:EnKF_results_coarse_fine}
\end{figure}

The first two DA runs are performed using a classical EnKF, where all the ensemble members are run using the same CFD solver and numerical grid. They will be referred to as EnKF-CG (where the model uses the Coarse grid for simulations) and EnKF-FG (the Fine grid is here used). Both DA applications rely on $N_e=39$ ensemble members. This choice is in agreement with previous DA analyses in the literature using CFD models \cite{Moldovan2024_cf} to provide a reasonably good approximation for the state covariance, but also to a reference benchmark to evaluate computational costs of the DA techniques. The ensemble members are initialized with random values of $\alpha \in [17^\circ, \, 19^\circ]$. The ensemble members are run simultaneously on 39 Intel(R) Xeon(R) Gold 5218R CPU cores, with one CPU core dedicated to each ensemble member. Additionally, the EnKF algorithm itself is executed on a single identical CPU core. The evolution of the DA optimized value for $\alpha$ is shown in \autoref{fig:EnKF_results_coarse_fine}, for both EnKF-CG and EnKF-FG runs. A rate of convergence parameters $\epsilon_\alpha$ is introduced and shown, which mathematical definition is:
\begin{equation}
    \epsilon_\alpha = \left\langle \left\Vert\frac{\alpha_{k-1} - \alpha_k}{\alpha_k} \right\Vert \right \rangle_{10} \label{eq:convergence_parameter_EnKF}
\end{equation}
Here $\alpha_{k-1}$ represents the updated angle of attack of the previous EnKF analysis and $\alpha_k$ is the optimized angle of attack of the current EnKF iteration. The moving average is defined as $\left \langle \cdot \right \rangle_{10}$ with an averaging window of values from ten consecutive analysis steps. 
The convergence criterion fixed for the EnKF analysis is selected for $\epsilon_\alpha < 10^{-4}$. 
First of all, both DA procedures reach convergence towards a suitable value of $\alpha$. More precisely, the value $\alpha = 25.45^\circ$ is obtained in around 100 analysis phases for the run EnKF-FG, while the optimized angle of attack for the run EnKF-CG converges to $\alpha = 25.33^\circ$ after around 240 analysis steps. While the values of $\alpha$ obtained for the two DA runs are not exactly equal to the target $\alpha = 25^\circ$, they are however consistent with the uncertainty in the measurements of the sensors that has been selected for the DA procedure. Higher precision can be obtained using inflation or reducing the uncertainty of the observation, but this point is not investigated here for sake of conciseness. Therefore, these results indicate that the usage of the Coarse grid requires a larger number of DA iterations to reach convergence. This point is not surprising, as the Fine grid used for the run EnKF-FG is significantly closer to the resolution level of the Reference grid used to produce the observation. The closer resemblance of the flow snapshots between model and observation leads to a faster convergence. However, this favorable feature is associated with a significant increase in computational costs due to the finer grid used. \autoref{tab:computational_costs_all} provides a detailed estimate of the computational costs necessary to perform the EnKF analyses presented in this section. 
\begin{table}[h!]
\centering
\small
    \begin{booktabs}{lccc}
    \toprule[1.5pt]
        \textbf{DA method} & \textbf{DA iteration number} & \textbf{CPUh} & \textbf{Total CPUh} \\ \midrule
        EnKF-CG        &    $240$      &     $0.42$     &  $100.8$   \\
        EnKF-FG        &    $86$       &    $10.42$      &   $896.12$  \\ \midrule[dashed]
        MGEnKF             &     $332$     &    $1.05$      &   $348.6$  \\
        MFEnKF             &      $345$    &     $1.09$     &  $376.05$   \\ \midrule[dashed]
        ConvMGEnKF        &     $102$     &      $2.73$    &   $278.46$  \\
        ConvMFEnKF        &     $112$     &    $3.1$      &  $347.2$   \\ \bottomrule[1.5pt]
    \end{booktabs}
    \caption{Computational cost for the DA runs performed.} \label{tab:computational_costs_all}
\end{table}
While the EnKF-CG run average cost is 0.42 CPUh to perform one complete cycle of EnKF forecast and analysis, 240 iterations are necessary to achieve a correct convergence for the optimized parameter, which leads to total computational cost of nearly 100 CPUh. On the other hand, the EnKF-FG run demands, on average, a computational load of 10.42 CPUh for a single cycle of forecast and analysis, whereas 86 cycles are needed to reach convergence, for a total of 900 CPUh. One can conclude that performing the EnKF using the Fine grid requires around nine times more resources to achieve a similar result than when using the Coarse grid.

\begin{figure}[h!]
    \centering
    \begin{tabular}{cc}
       \includegraphics[width=0.45\linewidth]{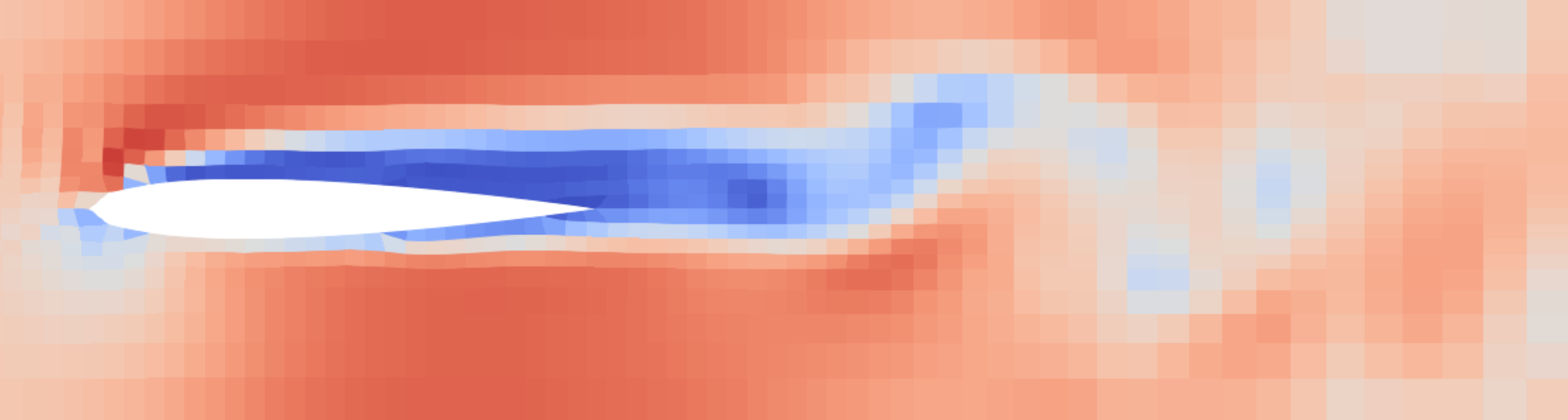}  &  \includegraphics[width=0.45\linewidth]{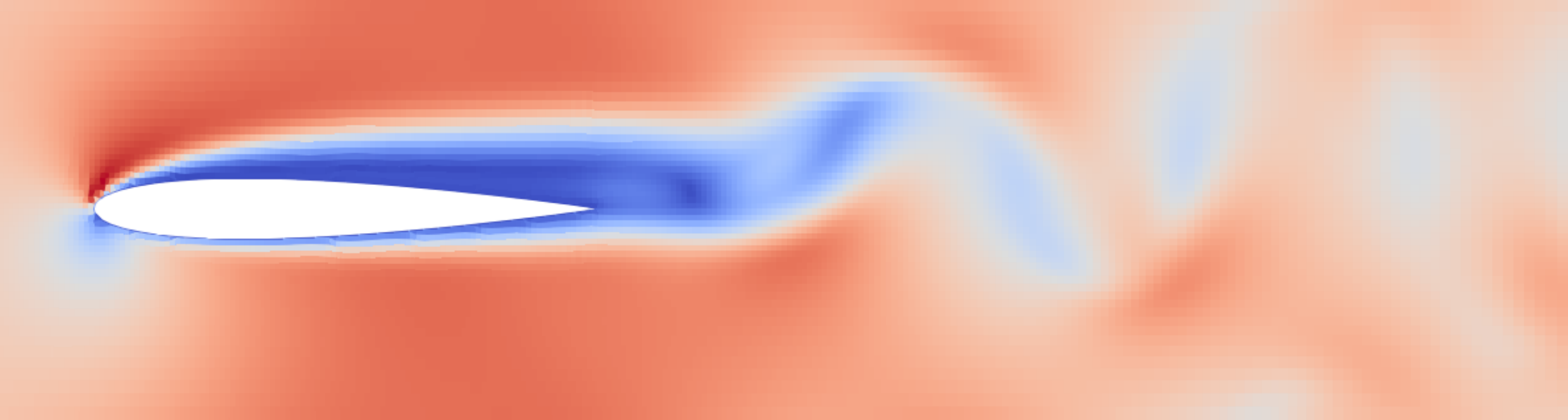} \\
       {\small (a)} & {\small (b)} \\[0.2cm]
       \includegraphics[width=0.45\linewidth]{figures/colorbar_0.0-0.82_UMag_over_a.pdf} & \includegraphics[width=0.45\textwidth]{figures/reference_mesh_U_instant.pdf} \\
        & {\small (c)}
    \end{tabular}
    \caption{Isocontours of the instantaneous normalized velocity magnitude $\Vert {\mathbf{u}} \Vert / a_\infty$ calculated at $t=25$ for (a) EnKF-CG run ($\alpha=25.33^\circ$); (b) EnKF-FG run ($\alpha=25.45^\circ$); (c) Reference simulation ($\alpha=25^\circ$).}
    \label{fig:init_EnKF_coarse_fine_grids}
\end{figure}

\begin{figure}[h!]
    \centering
    \begin{tabular}{cc}
        \includegraphics[width=0.45\linewidth]{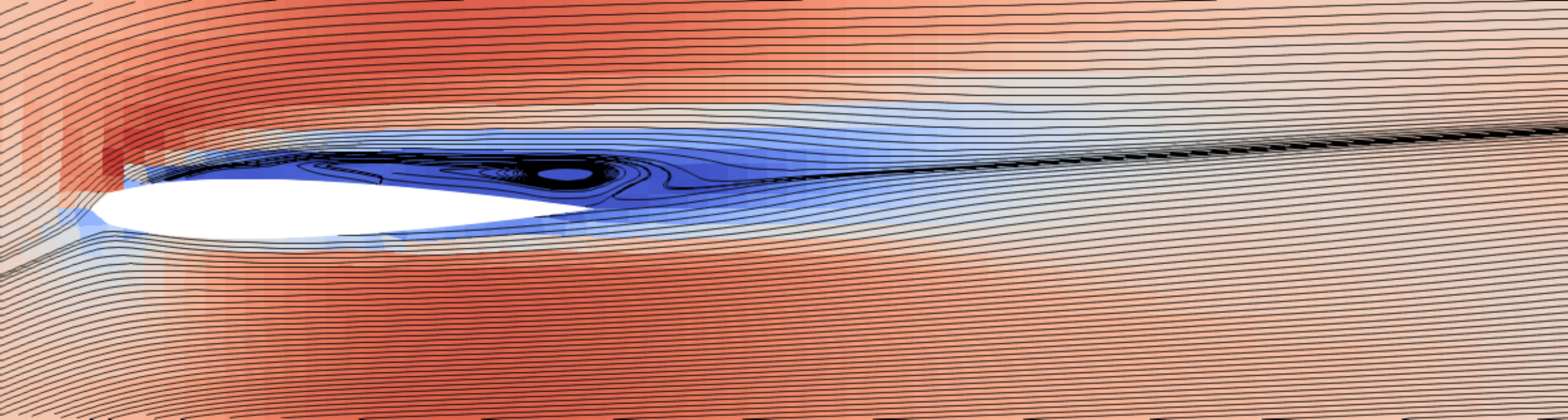} & \includegraphics[width=0.45\linewidth]{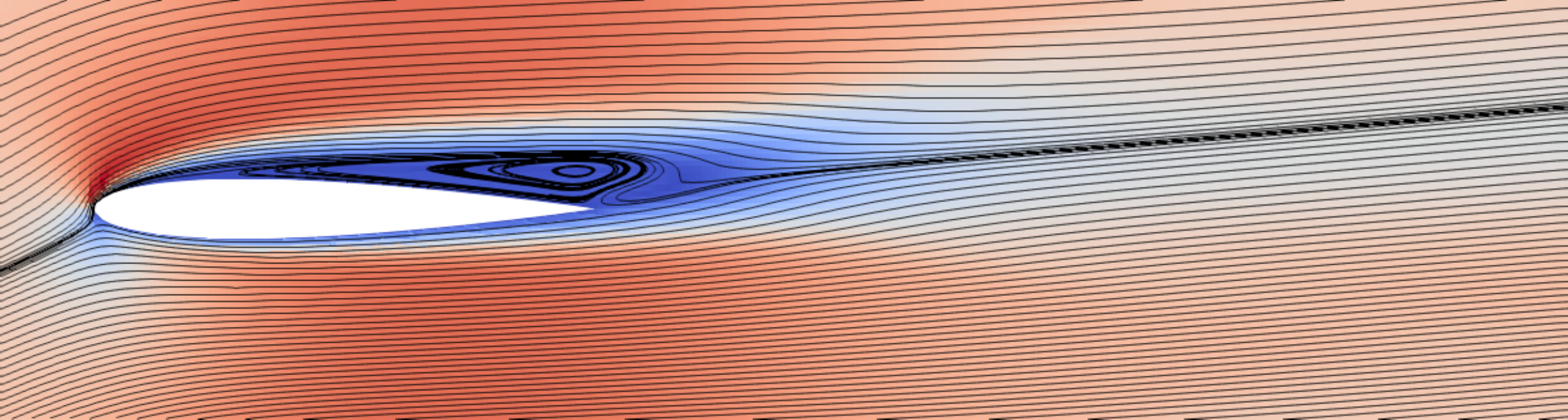} \\
        {\small (a)} & {\small (b)} \\[0.25cm]
        \includegraphics[width=0.45\linewidth]{figures/colorbar_0.0-0.82_UMeanMag_over_a.pdf} & \includegraphics[width=0.45\linewidth]{figures/reference_mesh_streamlines_UMean_bckgrd.pdf} \\
            & {\small (c)}
    \end{tabular}
    \caption{Isocontours of the time-averaged normalized velocity magnitude $\Vert \overline{\mathbf{u}} \Vert / a_\infty$ calculated at $t=25$ for (a) EnKF-CG run ($\alpha=25.33^\circ$); (b) EnKF-FG run ($\alpha=25.45^\circ$); (c) Reference simulation ($\alpha=25^\circ$).}
    \label{fig:streamlines_DA_coarse_medium}
\end{figure}

The physical features of the flow predicted  by the DA runs are now investigated. The isocontours of the instantaneous normalized velocity magnitude $\Vert \overline{\mathbf{u}} \Vert / a_\infty$ fields are shown in \autoref{fig:init_EnKF_coarse_fine_grids} and compared with the reference simulation for the time $t=25$. 
Even if the optimized angle of attack $\alpha$ converged to a satisfying value for both EnKF runs, the instantaneous velocity field predicted by the EnKF-CG exhibit visible discrepancies when compared with the reference simulation, in particular in the wake region. Similar conclusions can be drawn observing the time-averaged magnitude velocity profiles presented in \autoref{fig:streamlines_DA_coarse_medium}. This lack of resolution is responsible for the errors observed in the prediction of the bulk flow quantities, as shown in \autoref{tab:prior_DA_CL_CD_St_MGEnKF_MFEnKF_CNNs}. The results presented include the time average and standard deviation of the lift coefficient $C_L$, the drag coefficient $C_D$, the Strouhal number $St$ as well as the parameters $\Delta x_r/c$, $\Delta y_r/c$ and $x_{\textrm{sep}}/c$ measuring the size of the recirculation region. Comparison with the CFD run for $\alpha=25^\circ$ presented in \autoref{sec::2.3} shows that the results obtained by the DA run EnKF-CG for the bulk flow quantities are the same than the ones obtained with the simulation using the Coarse grid. This result is expected, because the state update via DA is performed only every $13.6 \, t_A$ and therefore the updated solution is periodically transported downstream without affecting statistics. Considering the case EnKF-FG, one can see in \autoref{fig:init_EnKF_coarse_fine_grids}  and \autoref{fig:streamlines_DA_coarse_medium} that the flow topology is well represented and qualitatively very similar to the reference simulation. This statement is true for both instantaneous and time-averaged flow solutions. Again for this case, one can see in \autoref{tab:prior_DA_CL_CD_St_MGEnKF_MFEnKF_CNNs} that the bulk flow quantities are almost identical to the ones provided for the CFD simulation for $\alpha=25^\circ$ using the Fine grid in \autoref{sec::2.3}. However, for this case discrepancies with the results for the reference simulation are significantly smaller. In both cases the DA algorithm is sufficiently accurate to identify a suitable angle of attack even if the flow snapshots do not exhibit a strong similarity. However, the DA implementation used for the runs EnKF-FG and EnKF-CG is not able to improve the underlying model prediction. One could argue that, if observation was available over shorter times windows, state estimation obtained from the DA process on both the Coarse grid and the Fine grid could improve the statistical quantities of the flow \cite{Meldi-DLES,valero_cf_2025}. However, the sampling frequency in real systems has limitations and its characteristic time is significantly larger than the usual size of time steps in numerical simulation \cite{Meldi2017_jcp}. Another possibility is to include subgrid corrective terms to improve the accuracy of the model. This task will be explicitly addressed in \autoref{sec:EnKF_with_CNN}. 
\begin{table}[h!]
\centering
\scriptsize
    \begin{booktabs}{lccccccccc}
    \toprule[1.5pt]
         & $\alpha \, [\textrm{deg}]$ & $\overline{C_L}$ & $C_L^\prime$ & $\overline{C_D}$ & $C_D^\prime$ & $St$ & $\Delta x_r/c$ & $\Delta y_r/c$ & $x_{\textrm{sep}}/c$ \\ \cmidrule{2-10}
    EnKF-CG & $25.33$ & $0.39$ & $0.01$ & $0.35$ & $0.004$ & $1$  &   $0.79$    & $0.086$  &  $0.42$ \\   
    EnKF-FG   & $25.45$ & $0.37$ & $0.008$ & $0.3$ & $0.003$ & $1.07$ &   $0.98$   & $0.078$  &  $0.2$  \\ \midrule[dashed]
    MGEnKF & $25.37$ & $0.37$ & $0.008$ & $0.31$ & $0.003$ & $1.08$ &    $0.98$   & $0.079$  &   $0.2$ \\
    MFEnKF & $25.09$ & $0.37$ & $0.007$ & $0.30$ & $0.003$ & $1.09$ &    $0.97$   & $0.077$  &  $0.21$ \\ \midrule[dashed]
    ConvMGEnKF  & $25.09$ & $0.37$ & $0.008$ & $0.31$ & $0.003$ & $1.09$   &  $0.98$   & $0.078$ & $0.2$  \\
    ConvMFEnKF   & $24.97$ & $0.37$ & $0.007$ & $0.30$ & $0.003$ & $1.09$  &  $0.98$   & $0.078$ & $0.2$ \\ \midrule[dashed]
    Ground truth  & $25$ & $0.37$  & $0.008$ & $0.3$  & $0.004$ & $1.11$  &  $1.04$   & $0.08$  &  $0.165$    \\ \bottomrule[1.5pt]        
    \end{booktabs}
    \caption{Time-averaged lift coefficient $\overline{C_L}$, drag coefficient $\overline{C_D}$, Strouhal number $St$ and features of the recirculation area $\Delta x_r/c$, $\Delta y_r/c$, $x_{\textrm{sep}}/c$. Data is provided for the DA runs performed as well for the reference simulation used to obtain observation.} \label{tab:prior_DA_CL_CD_St_MGEnKF_MFEnKF_CNNs}
\end{table}

In summary, the EnKF sensitivity analysis performed via two runs using different grids shows the algorithm is systematically able to optimize the free parameters affecting the test case. However, the DA procedure is sensitive to the choice of the grid of the model when the rate of convergence of the optimization is considered. The improvement observed using the Fine grid comes at the expense of computational overhead. On the other hand, despite the degraded convergence, the analysis on a Coarse grid leads to acceptable results, while maintaining a low computational effort. 
In the next section, ensemble realizations using both the Fine grid and the Coarse grid will be combined in the framework of multi-fidelity and multi-level analyses, using the previously introduced techniques MGEnKF and MFEnKF.

\subsection{DA application using multi-level and multi-fidelity techniques} \label{subsec:multilevel_enkf}
Two additional DA runs are now performed using the MGEnKF and the MFEnKF techniques. In order to provide a suitable comparison with the runs EnKF-FG and EnKF-CG, a total number of 39 ensemble members is performed using the same machine used for the previous study. More precisely, for the MGEnKF 38 ensemble members are simulated on the Coarse grid with a single member performed on the Fine grid. For the MFEnKF two \textit{principal} members are run using the Fine grid, two \textit{control} members shadowing the \textit{principal} members are performed on the Coarse grid and 35 \textit{ancillary} members also calculated using the Coarse grid. The other parameters of the DA run, including the time window between successive analysis phases and initialization of the parameter $\alpha$ for the inlet conditions, are set following the same criteria used for the runs EnKF-FG and EnKF-CG. In addition, the term $\mathcal{C}$ in \autoref{eq:mgenkf_forecast} is set to zero for this analysis using the MGEnKF i.e. the inner loop is not performed.

\begin{figure}[h!]
    \centering
    \includegraphics[width=0.5\linewidth]{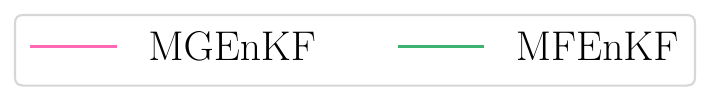}
    \includegraphics[width=0.65\linewidth]{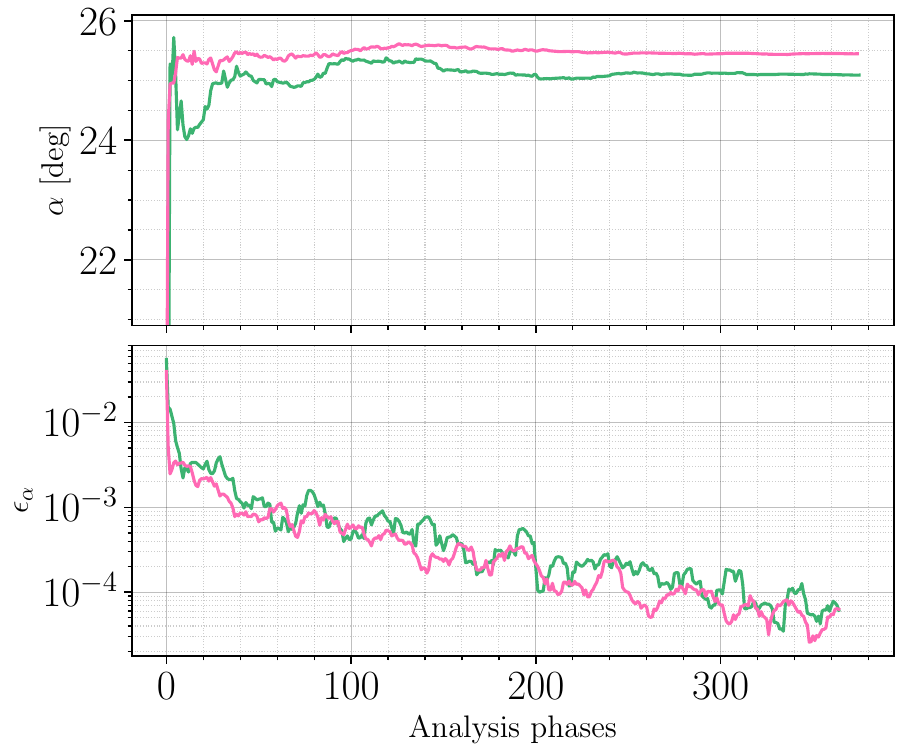}
    \caption{Top: evolution of the ensemble average angle of attack $\alpha$ applied to the inlet
boundary condition. Bottom: evolution of the convergence criterion $\epsilon_\alpha$.}
    \label{fig:MGEnKF_MFEnKF_alpha_epsilon_evolution}
\end{figure}

The evolution of the DA optimization for the parameter $\alpha$ is shown in \autoref{fig:MGEnKF_MFEnKF_alpha_epsilon_evolution}, with the coefficient $\epsilon_\alpha$ measuring the rate of convergence of the parametric inference (see \autoref{eq:convergence_parameter_EnKF}). Results are reported for both the MGEnKF and MFEnKF runs. Similarly to the two DA runs previously analyzed, the angle of attack is correctly obtained with a value of $\alpha = 25.37^\circ$ for the MGEnKF and $\alpha = 25.09^\circ$ for the MFEnKF. The two approaches require 332 (MGEnKF) and 345 (MFEnKF) DA analysis phases to reach convergence. 

The computational costs required by the multi-level / multi-fidelity approaches are now compared with the results from runs EnKF-FG and EnKF-CG. A summary is provided in \autoref{tab:computational_costs_all}. A first information that can be deduced is that the MGEnKF and the MFEnKF require a larger number of analysis phases to obtain the required convergence level of the parameter $\alpha$. This observation may be related to the need to combine information from different numerical models, which requires additional iterations for smoothing and balancing.
However, in terms of costs for a cycle of forecast and analysis of the DA procedure, the MGEnKF sums up to an average 1.05\,CPUh, while the MFEnKF is very close with 1.09\,CPUh. These requirements are around two or three times more than the computational times calculated for the run EnKF-CG, but still ten times smaller than the estimation provided for the run EnKF-CG. If one considers the total CPU cost to perform the optimization of the parameter $\alpha$, then the EnKF-CG is the best of the four DA runs. However, it is reminded that this application is the only one entirely relying on the use of the Coarse grid, and that the global prediction of the bulk flow quantities was not satisfying. In the following, the features of the flow simulated by the runs MGEnKF and MFEnKF, which rely on both the Coarse grid and the Fine grid, are investigated.

\begin{figure}[h!]
    \centering
    \begin{tabular}{cc}
           \includegraphics[width=0.45\linewidth]{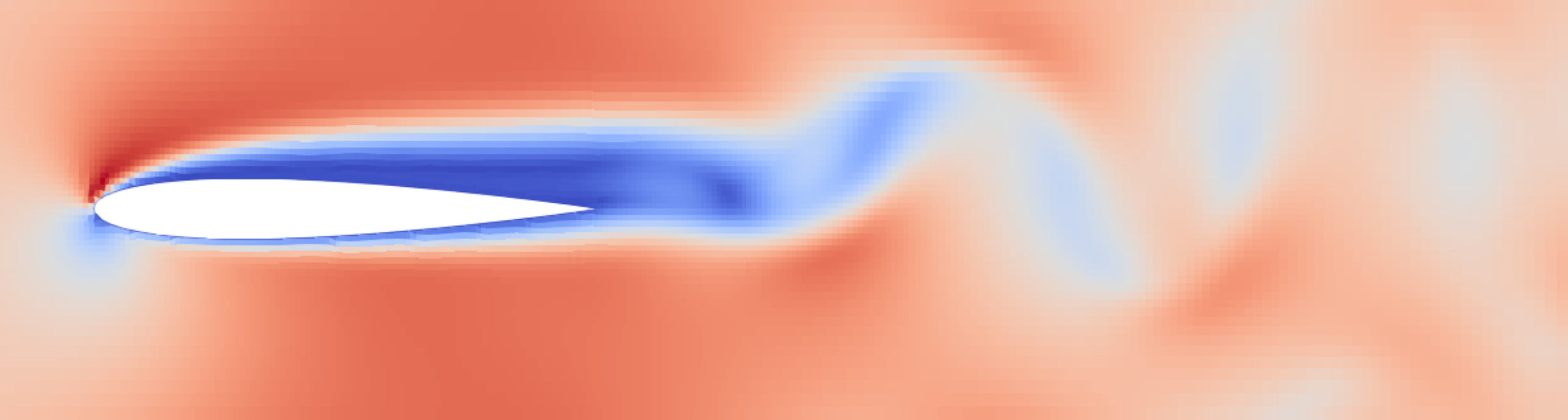}   &  \includegraphics[width=0.45\linewidth]{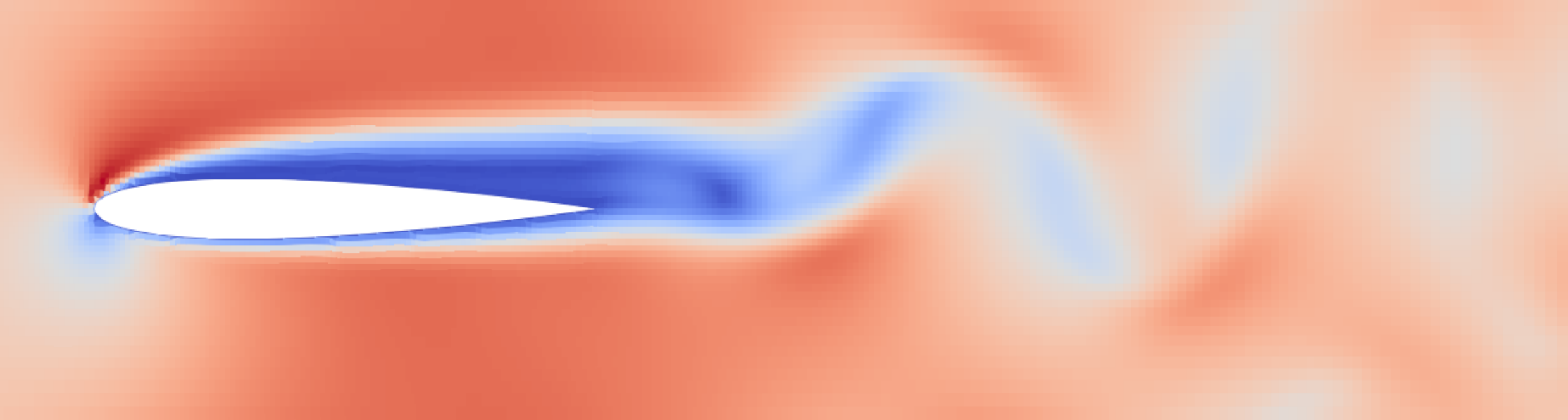} \\
        {\small (a)}   & {\small (b)} \\[0.2cm]
       \includegraphics[width=0.45\linewidth]{figures/colorbar_0.0-0.82_UMag_over_a.pdf}   & 
  \includegraphics[width=0.45\linewidth]{figures/reference_mesh_U_instant.pdf}     \\
          & {\small (c)} \\[0.25cm]
    \end{tabular}
    \caption{Instantaneous normalized velocity magnitude $\Vert \mathbf{u} \Vert / a_\infty$ at $t=25$ for (a)  MGEnKF optimized angle $\alpha=25.37^\circ$ on Fine grid; (b) MFEnKF optimized angle $\alpha=25.09^\circ$ on Fine grid; (c) Reference simulation ($\alpha=25^\circ$).}
    \label{fig:U_field_MGEnKF_MFEnKF_fine_grid}
\end{figure}
\begin{figure}[h!]
    \centering
    \begin{tabular}{cc}
       \includegraphics[width=0.45\linewidth]{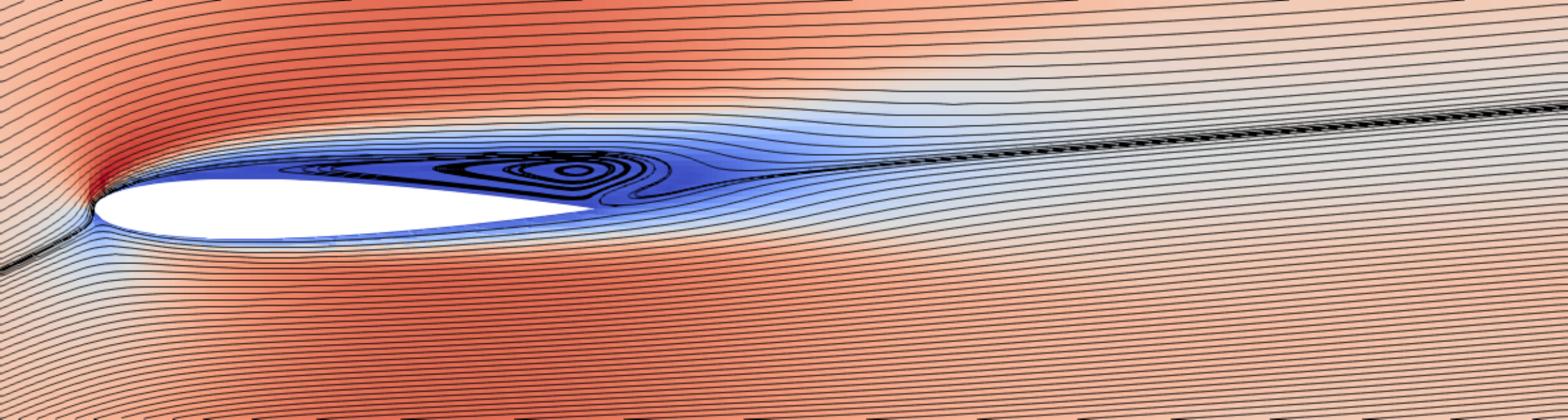}    & \includegraphics[width=0.45\linewidth]{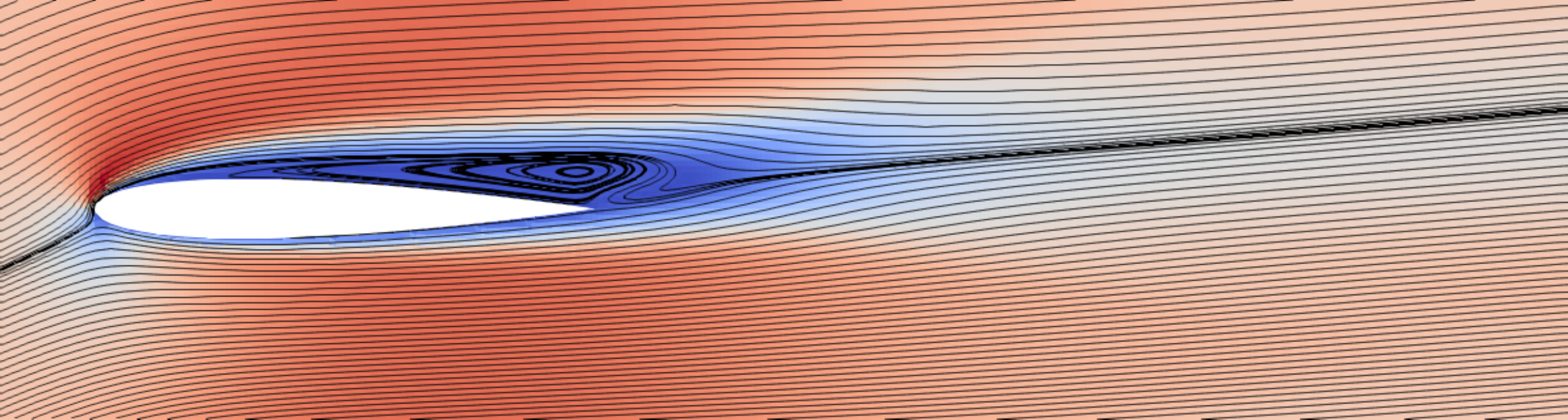} \\
       {\small (a)} & {\small (b)} \\[0.25cm]
      \includegraphics[width=0.45\linewidth]{figures/colorbar_0.0-0.82_UMeanMag_over_a.pdf}    &        \includegraphics[width=0.45\linewidth]{figures/reference_mesh_streamlines_UMean_bckgrd.pdf} \\
       & {\small (c)}
    \end{tabular}
    \caption{Isocontours of the time-averaged normalized velocity magnitude $\Vert \overline{\mathbf{u}} \Vert / a_\infty$ calculated at $t=25$ for (a) MGEnKF run ($\alpha=25.37^\circ$); (b) MFEnKF run ($\alpha=25.09^\circ$); (c) Reference simulation ($\alpha=25^\circ$).}
    \label{fig:streamlines_DA_MGEnKF_MFEnKF}
\end{figure}

\autoref{fig:U_field_MGEnKF_MFEnKF_fine_grid} shows the comparison of the normalized instantaneous velocity field $\Vert \mathbf{u} \Vert / a_\infty$ between the reference simulation and the runs MGEnKF and MFEnKF. Similarly to what was observed for the case EnKF-FG, the instantaneous flow field are extremely similar to the reference solution. The same conclusions can be drawn observing the time-averaged normalized velocity magnitude, which is shown in \autoref{fig:streamlines_DA_MGEnKF_MFEnKF}. The analysis of the bulk flow quantities reported in \autoref{tab:prior_DA_CL_CD_St_MGEnKF_MFEnKF_CNNs} also confirms that the statistical quantities calculated by the runs MGEnKF and MFEnKF are equivalent to the CFD realization for $\alpha=25^\circ$ using the Fine grid. Therefore, the question about the computational costs can be analyzed again now that more results about the accuracy in the prediction of the DA methods have been provided. If accurate prediction of the bulk quantities of the flow is required, the multi-level / multi-fidelity approaches clearly outshine the two strategies based on the usage of only one grid. MGEnKF and MFEnKF runs are significantly more precise than EnKF-CG realization and they have the same level of accuracy of EnKF-FG, but their total computational cost is between two and three times smaller.   

In conclusion, advanced multi-level and multi-fidelity methods have an edge over standard EnKF with an efficient blending of information obtained by high-resolution and low-resolution sources. However, one potential danger is represented by the fact that differences between the flow representation obtained by the Fine grid and the Coarse grid might be irreconcilable, leading to divergence of the filter. For the MGEnKF and MFEnKF, this can happen for example if during the forecast step the prediction of the members simulated on the Coarse grid exhibits a very different evolution from the prediction of the members run on the Fine grid, due to the cumulative effects of subgrid errors. This consideration suggests that it is crucial to obtain adequate accuracy on the low fidelity model while preserving low computational costs. In the next section, it will be investigated whether Machine Learning tools can enhance the accuracy of the low-fidelity solutions of ensemble members simulated on Coarse grids, thereby boosting the performance of the algorithm.

\clearpage

\section{CNN-enhanced multi-level strategies} \label{sec:EnKF_with_CNN}
In this section, tools from Machine Learning are used to reduce the discrepancy of the prediction of the flow states using the models based on the Coarse grid and the Fine grid. A first attempt to this task in the framework of the MGEnKF was proposed by Moldovan et al. \cite{moldovan_optimized_2022}. In the \textit{inner loop} a functional form was proposed for the term $\mathcal{C}$ in \autoref{eq:mgenkf_forecast} for the model run on the Coarse grid and its free coefficients were optimized using data streaming from the simulation on the Fine grid. While the procedure was successful, the main drawback is that a suitable functional form for the subgrid correction term $\mathcal{C}$ cannot be mathematically derived for complex flow cases. Therefore, computational costs can rapidly become prohibitive with the increase in the number of free parameters required to obtain an efficient description of such term. To this purpose, deep learning tools are here used in place of a functional form to represent the term $\mathcal{C}$. These data-hungry algorithms are here more easily trained thanks to the production of complete flow snapshot via EnKF. The complementary features of the DA and ML approaches here used are combined using the following scheme: 
\begin{itemize}
\item A first DA run is produced, such as the ones presented in \autoref{subsec:multilevel_enkf}. Data is stored at several times, in particular during the very first DA analyses where values of the parameter $\alpha$ are spread out for the different ensemble members.
\item The data is used to train a convolutional neural network (CNN) model which mimics the term $\mathcal{C}$. This black box tool is then integrated within the numerical simulations performed on the Coarse grid. 
\end{itemize}

\begin{figure}[h!]
    \centering
    \includegraphics[width=0.98\linewidth]{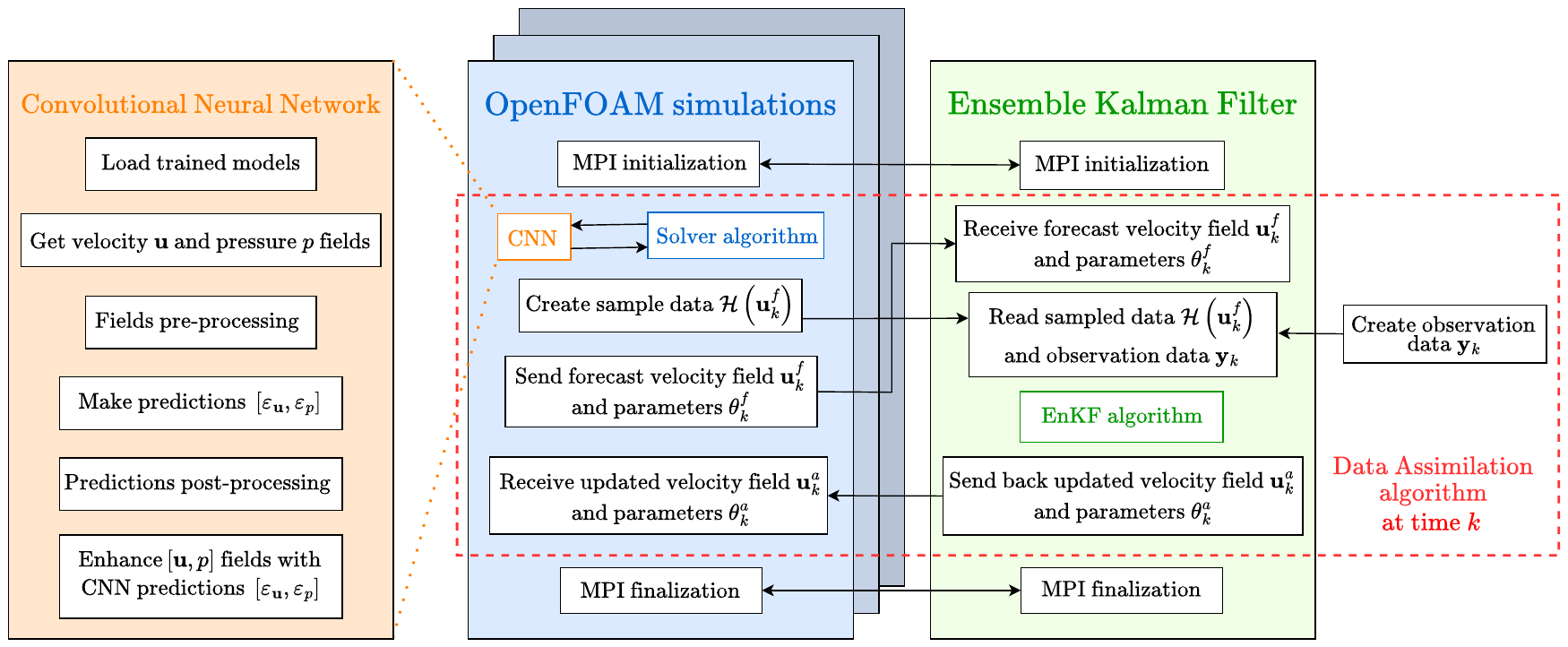}
    \caption{Schematic representation of CONES for online sequential Data Assimilation with CNN pre-trained tool integration.}
    \label{fig:CONES_with_CNN}
\end{figure}

Therefore, at a time step $k$, the CNN tool for the term $\mathcal{C}$ must provide a model correction:
\begin{equation}
    \mathcal{C}_k^{CNN} \left(\state{}_k^C\right) = \bm{\Phi} \left( \mathcal{M} \left( \state{}_k^F \right) \right) - \mathcal{M}^r \left( \state{}_k^C \right) \label{eq:cnn_output}
\end{equation}
in order to obtain $\bm{\Phi} \left( \state{}_k^F \right) = \state{}_k^C$ i.e. minimize the discrepancy in the flow prediction when using the Fine grid and the Coarse grid (see Equations \ref{eq:mgenkf_forecast} - \ref{eq:mgenkf_fine_grid_update}). It is here reminded that $\bm{\Phi}$ is the projection operator from the Fine grid to the Coarse grid. The optimization performed via the ML tool is only approximated as it is applied to the final state predicted by the CFD solver on the Coarse grid. This operation is not included in the recursive steps of the solver and therefore it does not necessarily comply with conservativity of the equations. However, thanks to the state updates jointly performed with the DA algorithms, it has been verified that state corrections are small and that effects associated with not conservativity are rapidly dissipated over a couple of iterations, similarly to initial condition effects. The choice to employ a direct state correction is here performed because of the simplicity in the implementation, which is connected with the operative features of the CNN algorithms working with images (states). However, more complex architectures where the ML forcing terms in the dynamic equations can be envisioned \cite{Villiers2025_ftc}, and they will be discussed in the conclusions.
The state corrections via ML tools is obtained with the generation of several distinct CNN models, one for each flow physical quantity. 
This architecture is formalized and implemented within the CONES library, adhering to the scheme presented in \autoref{fig:CONES_with_CNN}. This approach leverages the feature extraction capabilities of CNNs to address discrepancies arising from subgrid errors, thereby improving the predictive possibilities of the computationally less expensive simulations using the Coarse grid.


\subsection{CNN training}


The training process for the CNN models is now described. In Machine Learning, data quality is paramount for model performance. Suboptimal data can lead to significant degradation in training efficiency and, consequently, in the predictive capabilities of the resulting CNN. 

\begin{figure}[h!]
    \centering
    \includegraphics[width=0.92\linewidth]{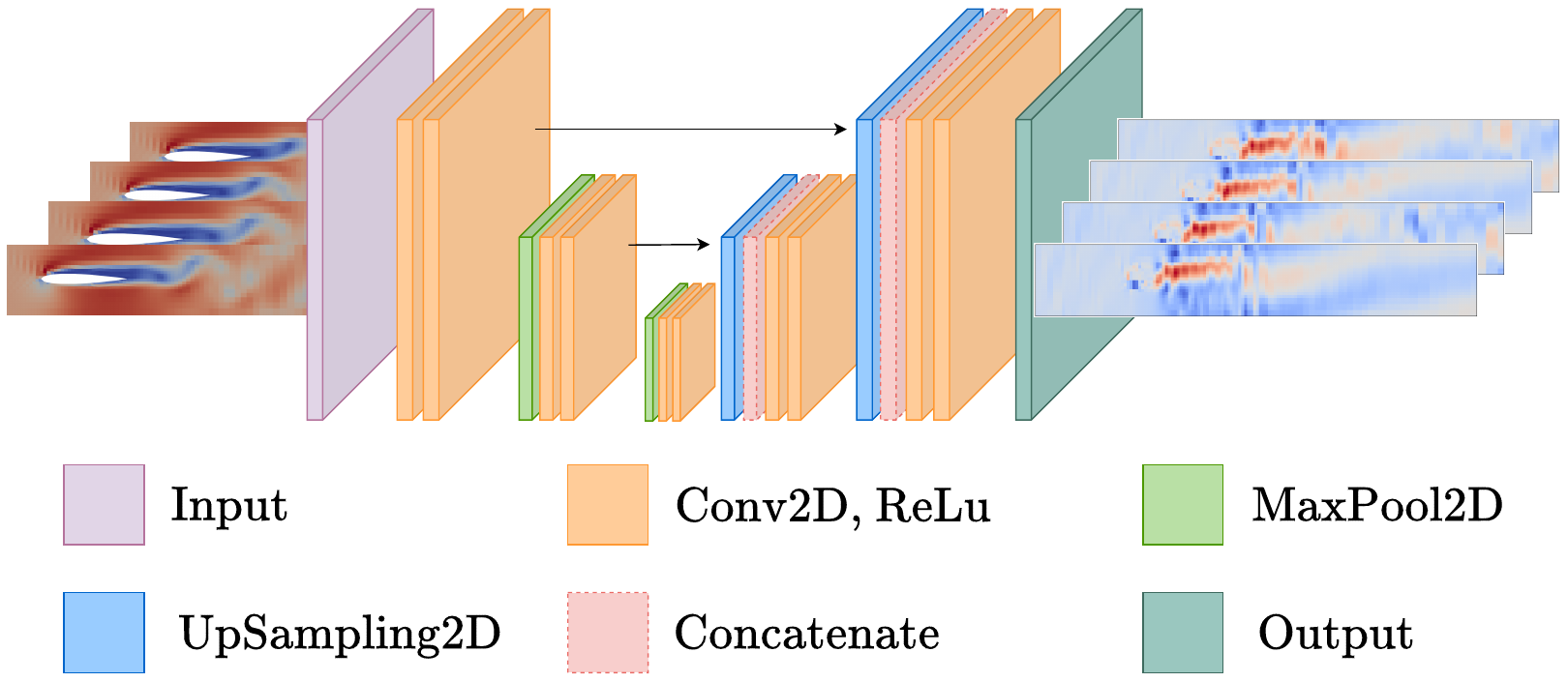}
    \caption{CNN architecture with input and corresponding output data.}
    \label{fig:CNN_input_output}
\end{figure}

The dataset used for training consists of $32 \, 000$ paired instances of input and output fields. Specifically, the input data corresponds to the flow fields acquired from the Coarse grid simulations, while the output data, according with \autoref{eq:cnn_output}, is the computed difference between the solution projected from the Fine grid and the input one. A visual representation is given in \autoref{fig:CNN_input_output}.

The aforementioned $32 \, 000$ input/output pairs were generated from a MFEnKF run. This simulation employed a specifically configured architecture consisting of 10 principal members, 10 control members and 10 ancillary members. This design choice is very important as it influences the diversity and representativeness of the generated data, which directly impacts the generalizability of the trained models.
The prior angle of attacks for this data generation are randomly set between 17 and 19 degrees, to be optimized towards values approaching 25 degrees as described in the previous studies.
Data acquisition from the DA cycle was performed rigorously across four distinct forecast/analysis phases, with measurements captured at every discrete time-step. This comprehensive sampling strategy resulted in a total of $4 \times 800 \times 10 = 32\, 000$ snapshots, ensuring a sufficiently large and varied dataset for robust model training. It is crucial to note that the input for each pair is exclusively derived from the control member data, providing a consistent baseline for the model's predictive task. In contrast, the output $\mathcal{C}^{CNN}$, is a computed quantity derived from both principal and control members. One can also see that, during the four DA analyses, the training procedure has access to data for different angles of attack $\alpha$, therefore enriching the information provided to the ML algorithm.


The presented case study employs a body-fitted mesh, meaning the  flow fields extracted from OpenFOAM are not in a format suitable for CNNs. CNN and neural networks usually perform matrix operations, which unstructured grids do not inherently provide. To address this issue, a grid projection strategy is used. The flow fields from the CFD solver are projected onto a 2D-structured grid. This structured grid aligns with the computational domain dimension. Specifically, the Coarse-grid mesh, originally composed by $2\,600$ cells, is mapped to a $150 \times 24$ 2D-structured grid, thus resulting in $3\,600$ structured cells. This grid projection is performed using the \texttt{griddata} function within the Python Scipy library. Three distinct models, grouped into two functional families are trained:
\begin{enumerate}
    \item \textbf{Velocity Model (x-component)}: this model is trained to predict the term $\mathcal{C}^{CNN} _{u_x}$. 
    \item \textbf{Velocity Model (y-component)}: this model is trained to predict the term $\mathcal{C}^{CNN} _{u_y}$. 
    \item \textbf{Pressure Model}: this model is trained for to predict the term $\mathcal{C}^{CNN}_{p}$. 
\end{enumerate}
The three models are staggered in compliance to the state variable ordering as $\mathcal{C}^{CNN} _k$:

\begin{equation}\label{eq:cnn_model}
\mathcal{C}^{CNN} _k = \left( \mathcal{C}^{CNN} _{u_x}, \mathcal{C}^{CNN} _{u_y}, \mathcal{C}^{CNN} _{p}\right)
\end{equation}

and used to synchronize the predicted flow state so that \autoref{eq:mgenkf_forecast} and \autoref{eq:mfenkf_forecast}
now take the following form:

\begin{equation}\label{eq:mgenkf_cnn_forecast}
    \state{^\mathbf{C}}^f _{k} = \mathcal{M}\left(\state{}^f_{k-1}, \mathbf{\theta}_{k-1}\right) + \mathcal{C}^{CNN}_k \left(\state{^\mathbf{C}}^f_{k-1}\right)
\end{equation}

\begin{equation}\label{eq:mfenkf_cnn_forecast}
    \control{}_k = \mathcal{M}^\mathbf{\gamma}\left(\state{}^f_{k-1}, \mathbf{\theta}_{k-1}\right) + \mathcal{C}^{CNN}_k \left(\principal{}_{k-1}\right)
\end{equation}

The models for \autoref{eq:mgenkf_cnn_forecast} and \autoref{eq:mfenkf_cnn_forecast} are compiled using an \textit{Adam} optimizer \cite{jais_adam_2019} while the pressure model in \autoref{eq:cnn_model} employs the \textit{RMSProp} optimizer \cite{hinton_neural_2012}. Both \textit{RMSProp} and \textit{Adam} are adaptive learning rate optimizers extensively used in deep learning works. Their utility stems from their ability to autonomously adjust the learning rate for individual network parameters (weights and biases) during training, in contrast to a globally fixed learning rate. This adaptive characteristic significantly enhances their efficiency in navigating complex loss landscapes.
To mitigate overfitting and enhance generalization, the velocity models incorporate dropout layers \cite{wu_towards_2015} with a dropout rate of 10\,\%, strategically positioned subsequent to the activation layers. Model generalization is further assessed using a $k-$fold cross-validation approach \cite{stone_cross-validatory_1974} with $k=4$ folds. The dataset is partitioned such that 75\,\% is allocated for training and 25\,\% for testing purposes. Furthermore, a dedicated random batch of 50 input\,/\,output pairs was set aside from the primary dataset for post-training evaluation. All models are optimized using the Mean Squared Error (MSE) as loss function and are trained during 400 epochs. Training computational process leverages an Nvidia Rtx A6000 GPU, using Tensorflow and Keras Python parallel processing. The specific computational costs of the CNN training are itemized in \autoref{tab:CNN_training_costs}. 

\begin{table}[h!]
    \centering
    \begin{tabular}{lcc} \toprule[1.5pt]
     \textbf{Model}  &  \textbf{Preprocessing} (CPUh) &  \textbf{Training} (GPUh)  \\ \midrule
     Velocity models  & $0.3$  & $0.42$ \\
     Pressure model   & $0.3$ & $0.63$ \\ \bottomrule[1.5pt]
    \end{tabular}
    \caption{Computational costs of CNNs training process.}
    \label{tab:CNN_training_costs}
\end{table}

A first preprocessing step, performed on CPU, is essential for preparing the CNNs training data. This phase encompasses operation such as the interpolation of OpenFOAM fields from the unstructured to the structured grid. This step requires 0.3 CPUh. Subsequently, for the Machine Learning tool training stage are primarily borne by the GPU due to highly parallelizable nature of neural networks operations. The training of the velocity models required 0.42 GPUh, concurrently, 0.63 GPUh are demanded to train for pressure model.

\subsection{EnKF runs with CNN models}

\begin{figure}[h!]
    \centering
    \begin{tabular}{cc}
       \includegraphics[width=0.45\textwidth]{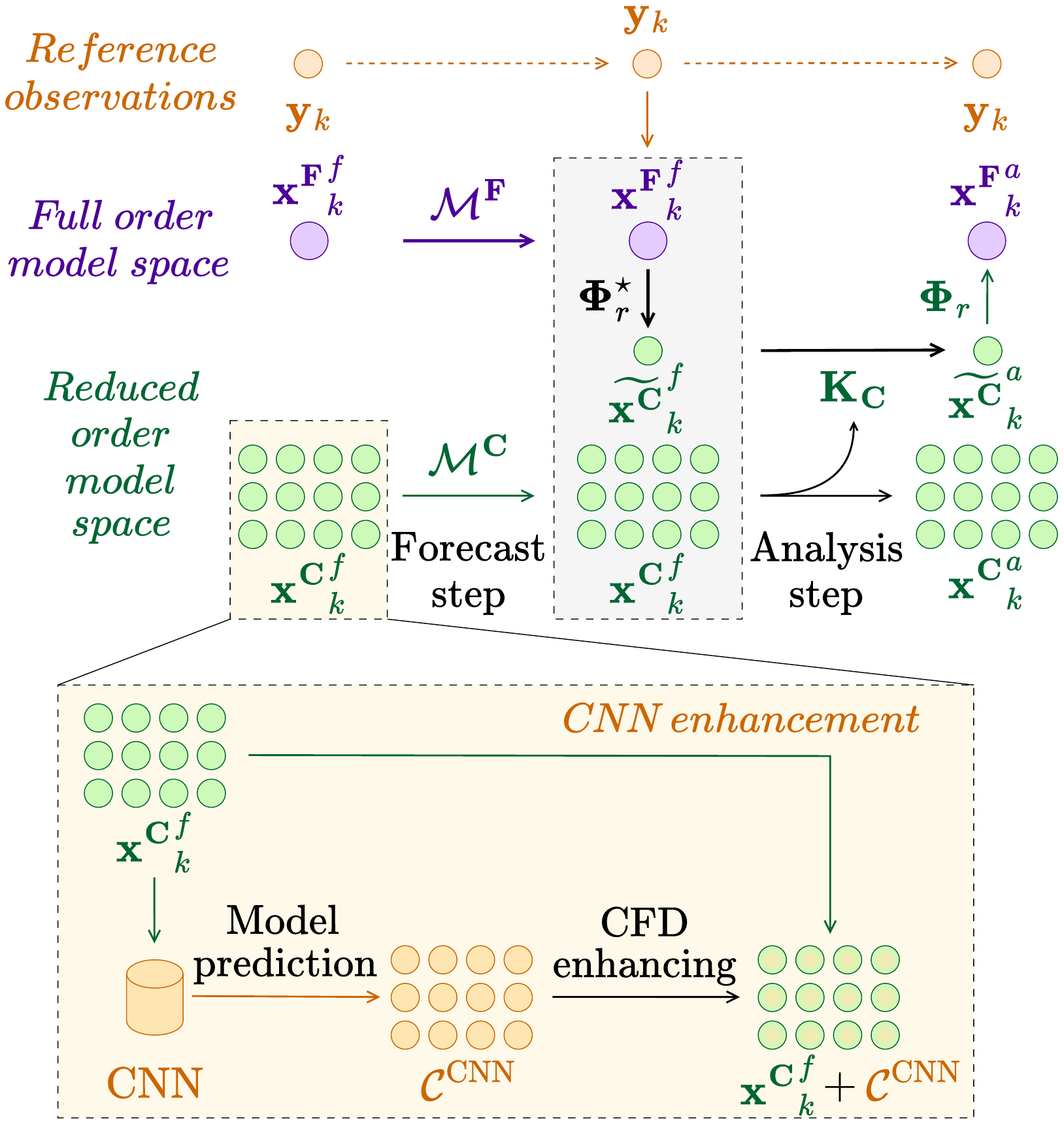}  & \includegraphics[width=0.45\textwidth]{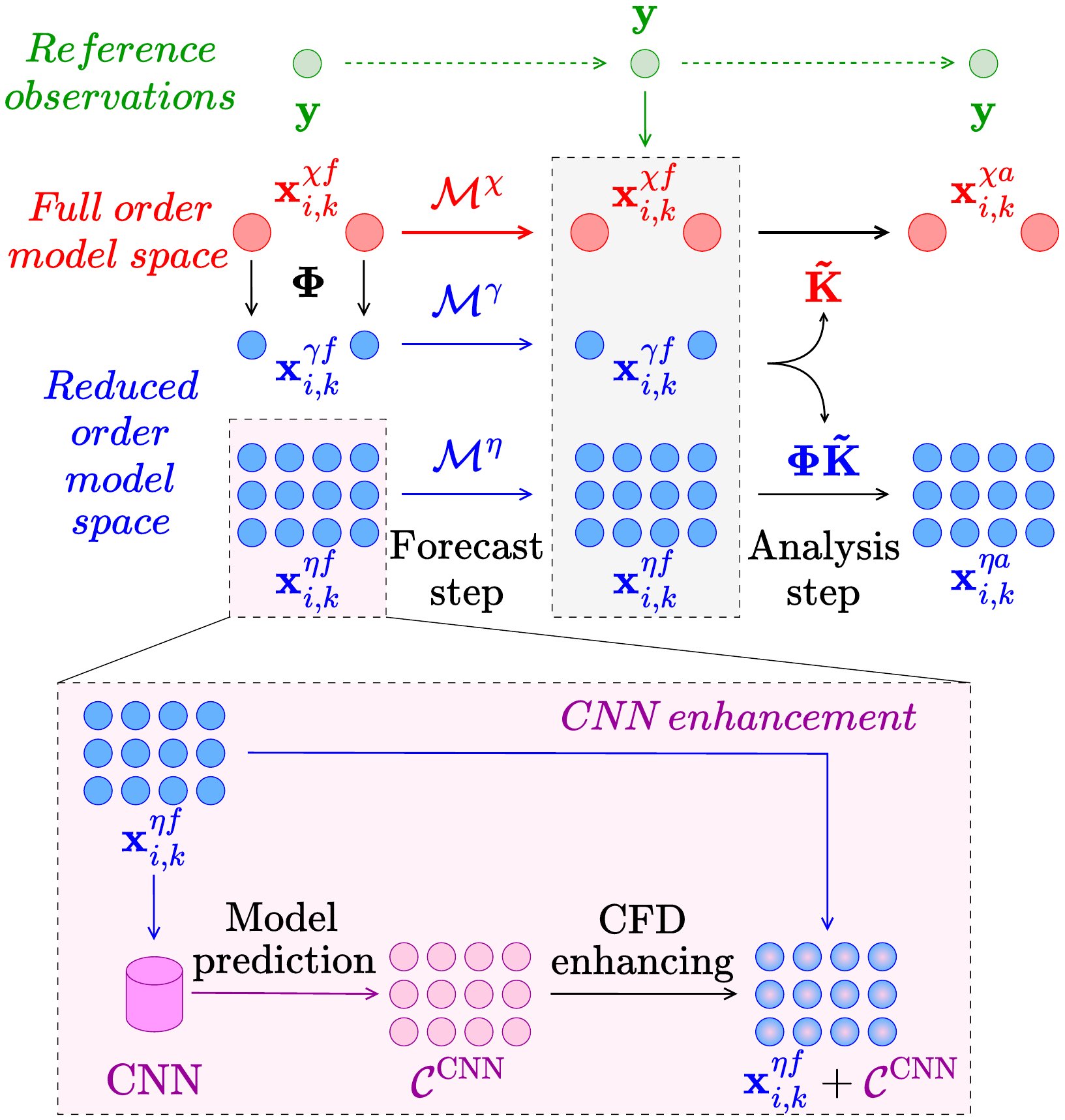} \\
       {\small (a)} & {\small (b)}
    \end{tabular}
    
    \caption{Scheme for the DA runs (a) ConvMGEnKF and (b) ConvMFEnKF, which are performed using the CNN tools used to improve the predictive capabilities of the models run on the Coarse grid.}
    \label{fig:ConvMG_ConvMF}
\end{figure}

The inclusion of the ML tools previously introduced with the DA framework is now described. In particular, two new DA runs are going to be performed. They exhibit the same set-up of the runs MGEnKF and MFEnKF but, in addition, the CNN tools are included to improve the accuracy of the prediction of the ensemble members run on the Coarse grid. These new runs will be referred to as ConvMGENKF and ConvMFEnKF, respectively 
and a schematic representation of the procedure is shown in \autoref{fig:ConvMG_ConvMF}.

\begin{figure}[h!]
    \centering
    \includegraphics[width=0.85\linewidth]{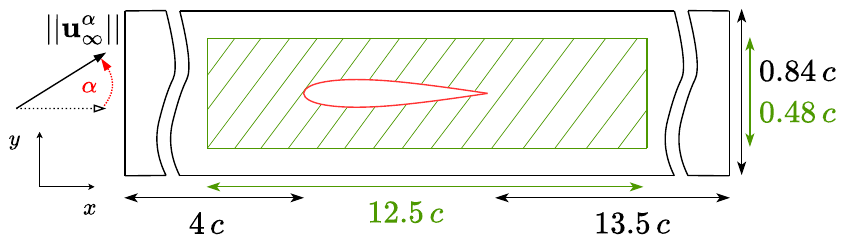}
    \caption{Region of application of the CNN tool for the CFD runs performed on the Coarse grid. The region is highlighted in green for improved visibility.}
    \label{fig:CNN_app_domain_on_airfoil}
\end{figure}

More details about the usage of the CNN tool are now provided. 
The parallel computing architecture used in CONES is the Multiple Program, Multiple Data (MPMD) \cite{gropp1994}. The MPMD structure is well suited for numerical problems where computational tasks have be handled by different algorithms (CFD solvers, Data Assimilation code and CNNs models) which need to interact and exchange data.
The inclusion of the pre-trained CNN models into CONES is direct through the usage of MPI communication routines. The CNN models are set-up to apply state corrections $\mathcal{C}^{CNN}_k$ to the flow field provided by the low fidelity models each $53$ time steps $\delta t$. The time window between successive applications of the CNN tool is chosen to be the equal to the average time for a shedding cycle, considering the value for the Strouhal number $St_{\textrm{ref}} = 1.11$ calculated for the reference simulation. Considering that DA analyses are performed every 800 time steps, that means that approximately $15$ state corrections via CNN are performed between consecutive applications of the EnKF. Sensitivity analyses have shown that a lower amount of state updates progressively degrades the accuracy of the solution, which finally becomes the original prediction of the model using the Coarse grid. On the other hand, more frequent state update are more expensive without significant improvement in accuracy.  
The application of the CNN tool is limited to a specific spatial subdomain of size $x/c \in [-2.5,10], \, y/c \in [-0.24,0.24]$, as shown in \autoref{fig:CNN_app_domain_on_airfoil}. The size has been selected to take into account the region in which the larger discrepancies in the state prediction are observed, which corresponds to the airfoil region. This choice is performed to reduce computational costs and to avoid potential spurious effects due to interaction with the boundary conditions. 


Results obtained for the two DA runs are now presented and discussed. 
\begin{figure}[h!]
    \centering
    \begin{tabular}{cc}
      \includegraphics[width=0.45\linewidth]{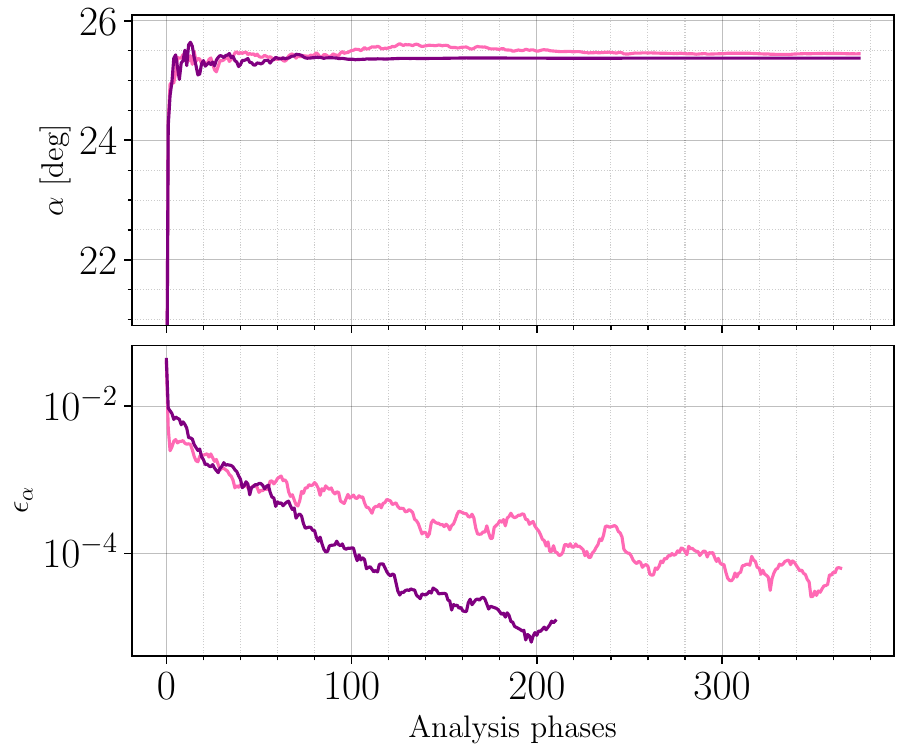}   & \includegraphics[width=0.45\linewidth]{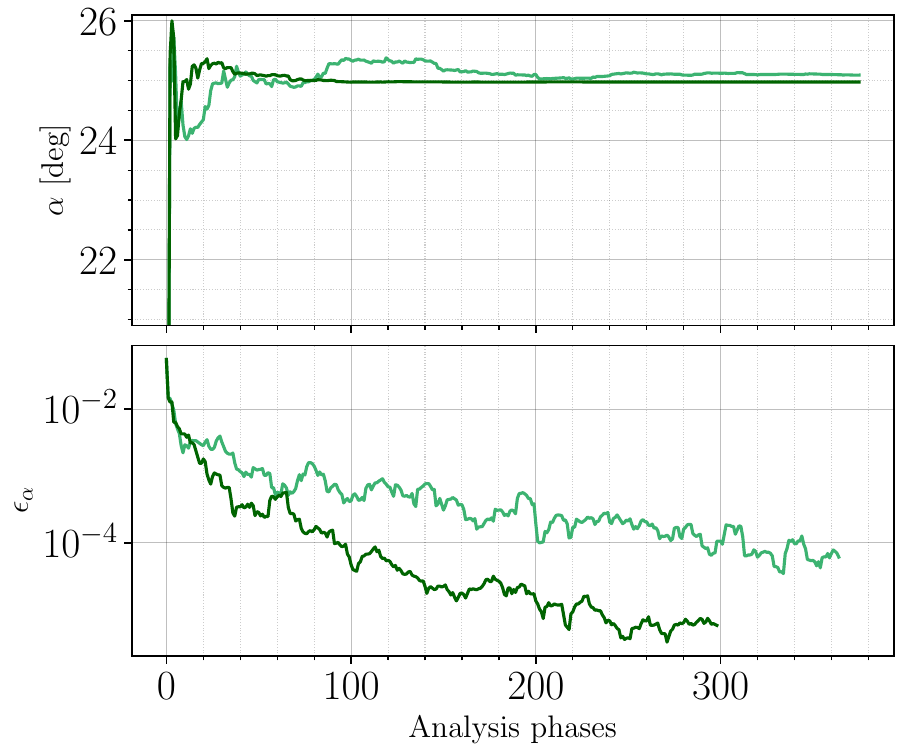} \\
      \includegraphics[width=0.45\linewidth]{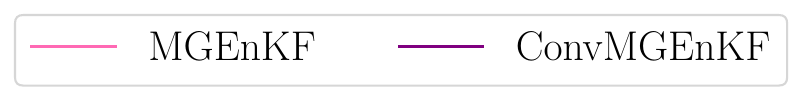}   & \includegraphics[width=0.45\linewidth]{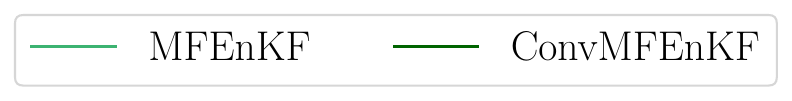}  \\  
      {\small (a)} & {\small (b)}
    \end{tabular} \\
    \caption{Top: Evolution of the angle of attack $\alpha$ of the inlet boundary condition. Bottom: Evolution of the convergence criteria $\epsilon_\alpha$. (a) Comparison of MGEnKF and ConvMGEnKF; (b) Comparison of MFEnKF and ConvMFEnKF.}
    \label{fig:MGEnKF_ConvMGEnKF-MFEnKF_ConvMFEnKF_comparison}
\end{figure}
\autoref{fig:MGEnKF_ConvMGEnKF-MFEnKF_ConvMFEnKF_comparison} shows that the CNN augmented DA runs ConvMGEnKF and ConvMFEnKF are able to obtain a precise estimation of the parameter $\alpha$, with very similar values to the realizations MGEnKF and MFEnKF. However, both augmented DA runs obtain a significantly faster optimization in terms of analysis phases required. As illustrated in \autoref{fig:MGEnKF_ConvMGEnKF-MFEnKF_ConvMFEnKF_comparison} (a),  the optimized value for the angle of attack $\alpha=25.24^\circ$ (respecting the convergence criterion $\epsilon_\alpha < 10 ^{-4}$ previously introduced)  is obtained after 102 DA analyses for the run ConvMGEnKF against the 300 iterations required for the run MGEnKF. 
This acceleration in the convergence of the algorithm is however associated with an increase in computational computational resources to run the CNN tool. A summary is reported in \autoref{tab:computational_costs_all}. One can see that the cost for a single forecast-analysis DA step is equal to 2.73 CPUh for the ConvMGEnKF run. This cost is higher than the one required by the run MGEnKF (1.05 CPUh) but lower than the one required by the run EnKF-FG (10.42 CPUh). Despite the increased cost per cycle when compared with the run MGEnKF, the substantial acceleration in convergence speed leads to a clear reduction in overall computational costs. Considering that 102 steps are required to reach converge in the optimization process, the total computational cost for this run is approximately 278.46 CPUh. Therefore, the DA run augmented with CNN exhibits a rate of convergence similar to the one observed for the EnKF run using only the model with the Fine grid, but its global computational costs are significantly smaller. For this application, the overall computation costs of the MGEnKF is reduced by almost 20\,\%. The trade-off between increased per-cycle cost and reduced total iteration is clearly favorable for this case.
Similar considerations can be performed comparing the run ConvMFEnKF with the run MFEnKF as shown in \autoref{fig:MGEnKF_ConvMGEnKF-MFEnKF_ConvMFEnKF_comparison} (b).
The integration of the CNN tool in this case permits to obtain an optimized value for $\alpha = 24.7^\circ$ within 112 DA iterations. This rapid convergence of the optimization process underlines the effectiveness of the ML model also in the Multi-Fidelity framework.
The computational overhead per-cycle observed for this case is slightly increased when compared to the ConvMGEnKF one. In this case, the overall computational costs is reduced by approximately $7\,\%$ when compared to the MFEnKF run.

\begin{figure}[h!]
    \centering
    \begin{tabular}{cc}
      \includegraphics[width=0.45\linewidth]{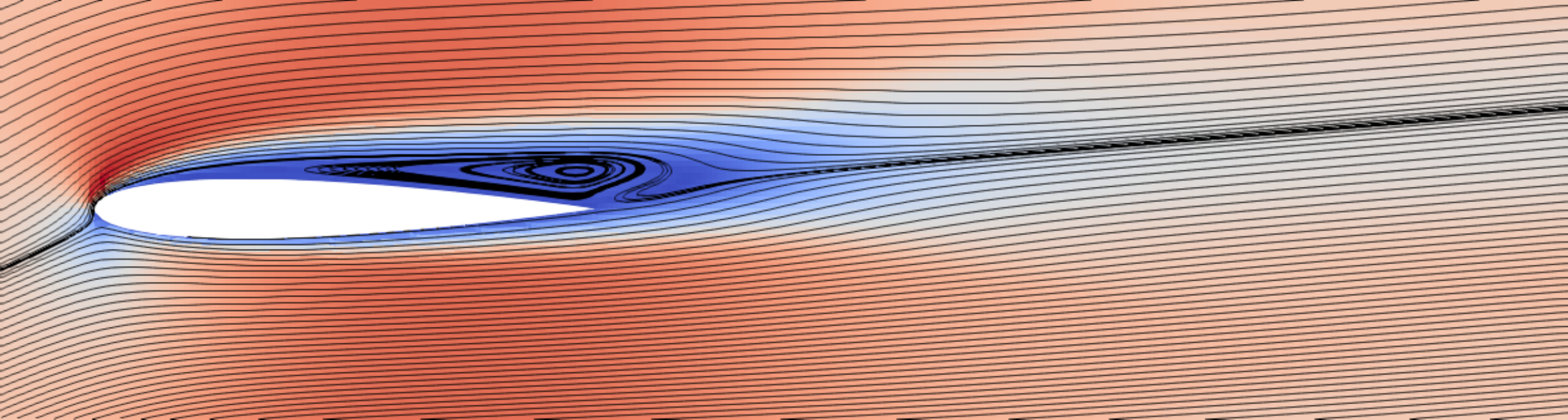}   & \includegraphics[width=0.45\linewidth]{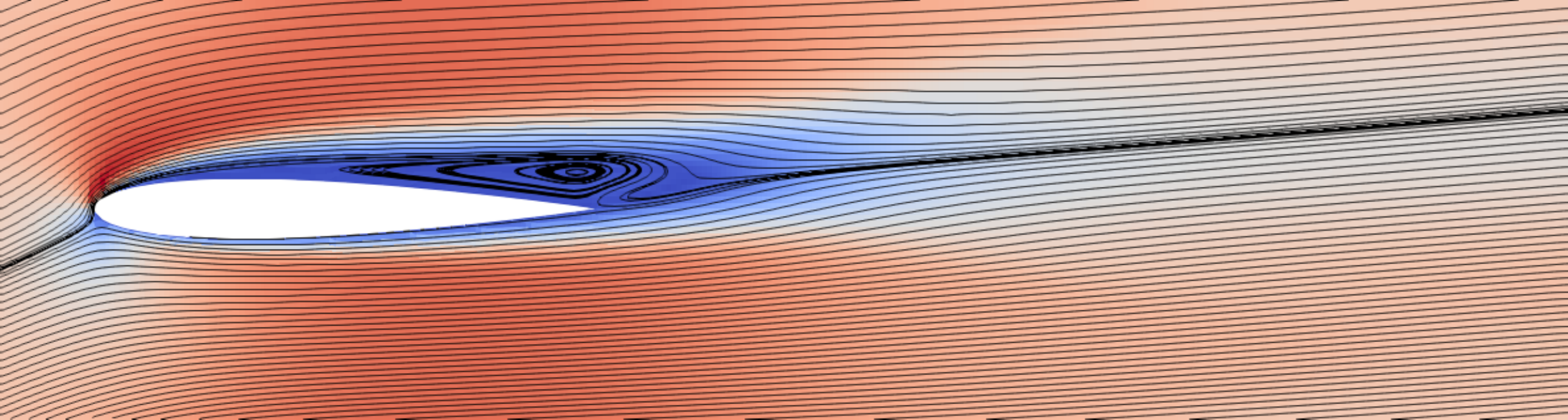} \\
      {\small (a)} & {\small (b)} \\[0.25cm]
      \includegraphics[width=0.45\linewidth]{figures/colorbar_0.0-0.82_UMeanMag_over_a.pdf}   &  \includegraphics[width=0.45\linewidth]{figures/reference_mesh_streamlines_UMean_bckgrd.pdf} \\
       & {\small (c)} 
    \end{tabular}
    \caption{Isocontours of the time-averaged normalized velocity magnitude $\Vert \overline{\mathbf{u}} \Vert / a_\infty$ calculated at $t=25$ for (a) ConvMGEnKF run ($\alpha=25.24^\circ$); (b) ConvMFEnKF run ($\alpha=24.97^\circ$); (c) Reference simulation ($\alpha=25^\circ$).
    }
    \label{fig:streamlines_CNNs}
\end{figure}

Finally, the analysis of the bulk flow quantities is presented in \autoref{tab:prior_DA_CL_CD_St_MGEnKF_MFEnKF_CNNs}. The results confirm that the runs ConvMGEnKF and ConvMFEnKF provide equivalent predictions for every quantity investigated to their counterparts MGEnKF and MFEnKF, but with diminished computational costs and higher efficacy in the optimization. Similar conclusions can be drawn by the analysis of the predicted flow fields, as shown in \autoref{fig:streamlines_CNNs} for the time-averaged velocity magnitude. Results from the CNN augmented DA runs exhibit very similar isocontours with the data from the reference simulation. 
Therefore, present results stress the potential to use to CNN tools to enhance CFD simulations and to represent subgrid effects. 


\section{Conclusion} \label{sec:conclusion}

The present study introduces a novel approach to enhance multi-level and multi-fidelity ensemble Data Assimilation techniques. To this purpose tools from Machine Learning, the Convolutional Neural Networks, are used to learn corrective terms to compensate subgrid errors due to the use of Coarse grids. This is done combining data from simulations available with different level of fidelity, whose prediction is given to the ML tools for training. The resulting tools target to reduce the discrepancy between high-fidelity and low-fidelity realizations, requiring only a moderate increase in computational resources. These techniques are validated for the simulation of a NACA\,0012 airfoil cascade at $Re= 1\,000$ and $Ma = 0.5$. This test case exhibits unsteady features which are controlled by the angle of the incidence $\alpha$ prescribed as inlet condition for the airfoil. Numerical simulations were performed using the solver \texttt{rhoPimpleFoam} of the library OpenFOAM. The Data Assimilation process targeted the optimization of the parameter $\alpha$, which was considered to be unknown. The calibration of such parameter was performed using data streaming from a reference simulation with $\alpha=25^\circ$, using observation from three sensors placed around the airfoil. Several DA runs have been performed using model runs on Fine and Coarse grids, as well as multi-level and multi-fidelity approaches. For the latter, investigations have also been performed including the model for CNN state correction. It was observed that the latter provides accurate results with a global decrease in the computational resources required, up to 20\,\%. In addition, all DA runs exhibited the accuracy level (in terms of prediction of bulk flow quantities and time-averaged features of the flow field) observed for simulations using the finest grid used in the EnKF procedure.

Future studies on this topic will be performed in order to improve the accuracy of the DA\,+\,ML tool and its versatility. First of all, the CNN tools used in this work have been trained via a preliminary DA investigation, and then applied as a black box model to a following run. One objective which will be pursued is an \textit{online} training, which will be performed during the DA procedure, and not before. While this point is clearly more challenging in terms of convergence of the algorithms, it is also less computationally expensive. A second point of future investigation deals with the feature of the CNN tool and how they integrate within the numerical CFD solvers. It was commented that present corrections are superposed to the solver prediction and, therefore, they might not be conservative with respect to the dynamic equations. This aspect could potentially lead to divergence of the calculation if the correction is important or if it is applied at every time step. One could envision to train the CNN tool to act as a forcing term in the equations, in order to provide more stability to the algorithm. A challenging point here is however associated with the recursive nature of the non-linear dynamic equations, which would require multiple evaluations of the CNN model during each time step. A last challenge deals with the accuracy of the DA\,+\,ML architecture. It was observed that the accuracy of the scheme was the same of the simulations run on the finest grid used. It was discussed that a lack of optimization of the model, as well as the time window between successive DA analyses, were responsible for this result. Future investigations could envision the training of a second CNN tool to improve the prediction of the model using the Fine grid. In this scenario, the target would be the state estimation produced by the EnKF.   

All of these points of discussion for future activities highlight the potential but also the need to explore approaches blending Data Assimilation and Machine Learning. These investigations are necessary to envision applications to cases of industrial interest such as transport engineering, urban flows or energy harvesting.

\section*{Acknowledgments}

This research work was carried out using TGCC storage and computing resources on the HPC Joliot Curie's Skylake partition, in the framework of the GENCI grant A0142A01741. The present work is funded by the French Ministry of the Armed Forces - Defense Innovation Agency (AID) and by the Hauts-de-France Region.

\bibliographystyle{elsarticle-num-names} 
\bibliography{cas-refs.bib}




\end{document}